\documentclass[aps,prl,twocolumn,superscriptaddress]{revtex4-2}
\usepackage{amsmath}
\usepackage{amsfonts}
\usepackage{amssymb}
\usepackage{lmodern,dsfont}
\usepackage{graphicx}
\usepackage[usenames,dvipsnames]{xcolor}
\usepackage{bm}
\usepackage{xr}
\usepackage[english=american]{csquotes}

\newcommand{\av}[1]{\langle #1 \rangle}
\newcommand{\Av}[1]{\left\langle #1 \right\rangle}
\newcommand{\nn}{\nonumber \\}
\newcommand{\n}{\nonumber}

\newcommand{\grad}{\bm{\nabla}}

\renewcommand{\eqref}[1]{Eq.~(\ref{#1})}

\begin{document}
	
	\author{Hiroyuki Odajima}
	\email{hiroyuki.odajima@yukawa.kyoto-u.ac.jp}
	\affiliation{Center for Gravitational Physics and Quantum Information, Yukawa Institute for Theoretical Physics, Kyoto University, Kitashirakawa Oiwakecho, Sakyo-ku, Kyoto 606-8502, Japan}
	\author{Andreas Dechant}
	\email{andreas.dechant@outlook.com}
	\affiliation{Department of Physics \#1, Kyoto University, Kitashirakawa Oiwakecho, Sakyo-ku, Kyoto 606-8502, Japan}
	\title{Measurement speed limit and Cram{\'e}r-Rao bound for time-averaged estimators}
	\date{\today}
	
	\begin{abstract}
		We consider a measured system $X$ containing a parameter $\theta$ of interest and the problem of estimating this parameter by observing a second, coupled measurement system $Y$.
		We show that, in thermodynamic equilibrium, the Fisher information of $Y$ associated with the parameter is bounded by that of $X$, implying that conservative interactions between $X$ and $Y$ reduce the precision of parameter estimation.
		Conversely, out of equilibrium, we show that the Fisher information of $Y$ can exceed that of $X$, so that non-conservative interactions can potentially increase the precision of parameter estimation.
		For idealized, back-action-free measurements, where the coupling to $Y$ does not perturb $X$, we derive a speed limit for the increase of Fisher information of $Y$.
		This speed limit only depends on the properties of the measured system $X$ and places a fundamental bound on the precision of any back-action-free measurement.
		As a corollary, we obtain a generalized Cram{\'e}r-Rao bound (CRB) for time-averaged estimators.
		We illustrate our results for linear dynamics and suggest a scheme for disturbance-free non-equilibrium sensing with improved accuracy.
	\end{abstract}
	
	\maketitle
	
	Estimating the values of parameters in a physical system from the measurement of observables is crucial for various experimental platforms.
	We may want to quantify the strength of the interactions between colloidal particles by observing their relative distance for example, or quantify the chemical potential difference driving a molecular motor by observing the motion of a tracer particle coupled to it.
	Since physical systems are often subject to considerable noise from their environment, the results of measurements are likewise random and parameters can only be estimated with finite precision.
	
	The fundamental limit of the precision of parameter estimation is governed by Fisher information \cite{Cover2012}.
	Suppose that the random configuration of the system $X$ is $\bm{x}$, and its statistics are described by a probability distribution $p^X_\theta(\bm{x})$ that depends on some parameter $\theta$ of interest.
	In order to determine $\theta$ from a measurement of $\bm{x}$, we use an estimator $\hat{\theta}(\bm{x})$, whose expected value is equal to the parameter $\av{\hat{\theta}} = \theta$.
	The Cram{\'e}r-Rao bound (CRB) \cite{RadhakrishnaRao1945, Cramer2016} provides the fundamental limit of the precision of any estimator,
	\begin{align}
		\text{Var}(\hat{\theta}) \geq \frac{1}{F^X_\theta} \label{cramer-rao} . 
	\end{align}
	Here, $\text{Var}(\hat{\theta}) = \av{\hat{\theta}^2} - \av{\hat{\theta}}^2$ is the variance of the estimator and $F^X_\theta$ is the Fisher information defined by
	\begin{align}
		F^X_\theta = \Av{\big(\partial_\theta \ln p^X_\theta \big)^2} \geq 0 \label{fisher-x}.
	\end{align}
	Thus, the Fisher information places a lower bound on the error of parameter estimation and provides a benchmark for constructing precise estimators with minimum variance.
	
	In many cases, in particular for small systems, we cannot observe the measured system $X$ directly, but rather its effect on another system $Y$, which we refer to as the measurement system.
	Physically, $Y$ may be a tracer particle bound to a molecular motor or a sensor that detects light scattered off a levitated particle.
	Since we can only directly access the configuration $\bm{y}$ of the measurement system $Y$, we must estimate the parameter $\theta$ from $\bm{y}$, that is, use an estimator $\hat{\theta}^Y(\bm{y})$ with $\av{\hat{\theta}^Y} = \theta$.
	Formally, the CRB \eqref{cramer-rao} also holds for $Y$,
	\begin{align}
		\text{Var}_\theta(\hat{\theta}^Y) \geq \frac{1}{F^Y_\theta} \label{cramer-rao-Y}, 
	\end{align}
	where $F^Y_\theta$ is defined analog to \eqref{fisher-x} using the probability distribution $p_\theta^Y(\bm{y})$ of $Y$.
	However, $F^Y_\theta$ generally depends in a complicated manner on both the properties of $Y$ and the nature of the coupling between $X$ and $Y$, which are often difficult to characterize precisely in practice.
	The central question addressed in this Letter is how we can characterize $F^Y_\theta$ using the properties of the measured system $X$. 
	This allows us, on the one hand, to formulate fundamental bounds on the precision of indirect parameter estimation and, on the other hand, to give a tangible benchmark for constructing precise estimators.
	
	\paragraph{Setup.}
	In this Letter, we focus on the overdamped Langevin equations \cite{Coffey2017} describing the joint dynamics of the measured system $X$ and the measurement system $Y$,
	\begin{subequations}
		\begin{align}
			\bm{\gamma}^X \dot{\bm{x}} &= \bm{f}^X_\theta(\bm{x}) + \bm{f}^{XY}(\bm{x},\bm{y})  + \sqrt{2 \bm{\gamma}^X T^X} \bm{\xi}^X \\
			\bm{\gamma}^Y \dot{\bm{y}} &= \bm{f}^Y(\bm{y}) + \bm{f}^{YX}(\bm{x},\bm{y}) + \sqrt{2 \bm{\gamma}^Y T^Y} \bm{\xi}^Y .
		\end{align} \label{langevin}%
	\end{subequations}
	$\bm{\gamma}^X$ and $\bm{\gamma}^Y$ are positive definite friction tensors, $T^X$ and $T^Y$ are the temperatures of the heat baths in contact with $X$ and $Y$ and $\bm{\xi}^X$ and $\bm{\xi}^Y$ are vectors of Gaussian white noises whose components are mutually independent.
	Let us remark on two important assumptions made in \eqref{langevin}: 
	First, the part of the force that depends on the parameter $\theta$, $\bm{f}^X_\theta$ only affects system $X$, while the remaining forces acting on $Y$ or mediating the interactions between $X$ and $Y$ are independent of $\theta$.
	Second, the noises acting on $X$ and $Y$ are independent; that is, any coupling between $X$ and $Y$ is caused by forces and not by environmental fluctuations.
	We stress that, as shown in the supplemental Material (SM) \cite{SM}, the results remain valid for underdamped dynamics; here, we focus on the overdamped case for simplicity.
	
	\paragraph{Equilibrium measurements.}
	We begin with the measurements in equilibrium states.
	In this scheme, we assume that $X$ and $Y$ are jointly in thermal equilibrium.
	In the context of \eqref{langevin} this implies that $X$ and $Y$ are in contact with a single equilibrium reservoir, $T^X = T^Y = T$,  and that the forces acting on the system are conservative, $\bm{f}^X_\theta = -\grad_x U_\theta^X$, $\bm{f}^Y = -\grad_y U^Y$, $\bm{f}^{XY} = -\grad_x U^\text{int}$ and $\bm{f}^{YX} = -\grad_y U^\text{int}$, where $U_\theta^X$ is the parameter-dependent potential energy of $X$, $U^Y$ is the potential energy of $Y$ and $U^\text{int}$ is the interaction energy.
	We imagine that, initially, $X$ and $Y$ are independently in equilibrium (i.~e.~$U^\text{int} = 0$), so that $Y$ carries no information about the parameter $\theta$, $F^Y_\theta(t = 0) = 0$.
	Next we introduce interactions between $X$ and $Y$ and let them relax to a new common equilibrium state.
	Due to the interaction, the final state of $Y$ generally depends on the parameter $\theta$, allowing us to infer it from a measurement of $Y$.
	Our first main result is that, in equilibrium, the relation
	\begin{align}
		F_\theta^{Y,\text{eq}} \leq F_\theta^{X,\text{eq}} \label{fisher-equilibrium}
	\end{align}
	holds.
	The proof of the relation \eqref{fisher-equilibrium} is given in Appendix A.
	This implies that, in thermal equilibrium, measuring $\theta$ using $Y$ is necessarily less precise than measuring $\theta$ from a direct observation of $X$.
	While this result may seem obvious, surprisingly, it only holds in equilibrium.
	
	\paragraph{Back-action free measurements.}
	Next, we introduce the back-action-free scheme, which is complementary to the equilibrium scheme.
	As before, initially, $X$ is in equilibrium and the state of $Y$ is independent of $\theta$.
	However, in contrast to before, we now assume that the interactions between $X$ and $Y$ do not affect $X$, while $Y$ depends on $X$, thus acquiring a dependence on $\theta$ due to the interaction.
	In \eqref{langevin}, this corresponds to $\bm{f}^X_\theta = -\grad_x U_\theta^X$ and $\bm{f}^{XY} = 0$ while $\bm{f}^{YX}$, $\bm{f}^Y$ and $T^Y$ are arbitrary.
	For simplicity, here we assume the friction tensor of $X$ is isotoropic; $\bm \gamma^X = \gamma^X \bm I$, where $\bm I$ is a unit matirix.
	Note that, the non-symmetric interaction necessarily corresponds to a non-conservative force, and thus the measurement corresponds to an out-of-equilibrium process.
	Our second main result is that, for back-action-free measurements, the Fisher information of $Y$ obeys the speed limit
	\begin{align}
		F_\theta^Y(t) \leq F_\theta^{X,\text{eq}} + \frac{1}{2} \mathcal{F}_\theta^{X} t \label{fisher-speed-limit} .
	\end{align}
	Here, we defined the rate $\mathcal{F}_\theta^{X}$; for a particle in a viscous thermal environment with friction coefficient $\gamma^X$ and temperature $T^X$, it is given by
	\begin{align}\label{fisher-rate}
		\mathcal{F}_\theta^{X} = \frac{1}{\gamma^XT^X} \Av{\big\Vert\partial_\theta \bm{f}^X_\theta\big\Vert^2 }.
	\end{align}
	We stress that $\mathcal{F}_\theta^{X}$ only depends on the properties of system $X$ and is independent of the choice of $Y$ and the details of the interaction, aside from the condition that there is no back-action.
	Therefore, \eqref{fisher-speed-limit} bounds the information that can be acquired about a parameter $\theta$ in the system $X$ by any measurement.
	The upper bound in \eqref{fisher-speed-limit} increases linearly with time, providing a speed limit on the rate at which Fisher information can increase. 
	Crucially, this suggests that the Fisher information of $Y$ can exceed the Fisher information of $X$, allowing for a more precise measurement of $\theta$. 
	We remark that while we have assumed that the friction tensor is isotropic (i.e., $\bm\gamma^X = \gamma^X\bm I$), however, our results can be extended straightforwardly for general fiction tensors (Appendix B).  
	
	\paragraph{Generalized CRB for time-averaged estimators.}
	It appears counter-intuitive that a measurement of $Y$ should provide a better estimate than a direct measurement of $X$.
	To illustrate this result, we consider the following concrete example. 
	If $\hat{\theta}(\bm{x})$ is an estimator of $\theta$ in $X$, we can define
	\begin{align}
		y(t) = \bar{\theta}(t) = \frac{1}{t} \int_0^t dt' \ \hat{\theta}(\bm{x}(t')) .
	\end{align}
	In this case, the function of the \enquote{measurement system} $Y$ is to perform a time-average of the estimator $\hat{\theta}$.
	Since this does not impact $X$, it corresponds to a back-action-free measurement.
	It is clear that we have $\av{y} = \av{\bar{\theta}} = \av{\hat{\theta}} = \theta$, so that $y = \bar{\theta}$ is likewise an estimator of $\theta$.
	However, time-averaging reduces the fluctuations of the estimator and thus the variance of $\bar{\theta}$ is not bounded by the Fisher information of $X$. 
	Instead, from \eqref{fisher-speed-limit} we obtain a generalized CRB for time-averaged estimators, which is our third main result;
	\begin{align}
		\text{Var}(\bar{\theta}) \geq \frac{1}{F_{\theta}^{X,\text{eq}} + \frac{1}{2} \mathcal{F}_\theta^{X} t} . \label{time-average-cramer-rao}
	\end{align}
	This provides a bound on the improvement in parameter estimation due to time-averaging.
	The reason why time-averaging can \enquote{beat} the normal CRB \eqref{cramer-rao} is that the latter only applies to estimators that depend on the instantaneous configuration of system $X$.
	By contrast, the time-averaged estimator also depends on past configurations of $X$; it is this additional information that allows us to obtain a more precise estimate of the parameter.

	\paragraph{Measurements in linear systems.}
	As a concrete, solvable example containing both measurement schemes, we consider the linear coupled Langevin equations
	\begin{align}
		\gamma^X \dot{x} &=  -k^X (x - \theta) + \kappa^{XY} y + \sqrt{2 \gamma^X T^X} \xi^X, \nn
		\gamma^Y \dot{y} &= -k^Y y + \kappa^{YX} x  + \sqrt{2 \gamma^Y T^Y} \xi^Y . \label{langevin-linear}%
	\end{align} 
	The system $X$ is an overdamped particle in a parabolic trapping potential $U^X_\theta = k^X (x-\theta)^2/2$ with spring constant $k^X$ located at $x = \theta$.
	Likewise, we can imagine $Y$ to be an overdamped particle in the trapping potential $U^Y = k^Y y^2/2$.
	The coupling constants $\kappa^{XY}$ and $\kappa^{YX}$ determine the interaction between $X$ and $Y$.
	If $\kappa^{XY} = \kappa^{YX}$, then the interactions derive from the interaction potential $U^\text{int} = \kappa^{XY} x y$, corresponding to conservative interactions.
	If $T^X = T^Y = T$, then the overall system is in thermal equilibrium.
	Conversely, if $\kappa^{XY} \neq \kappa^{YX}$ or $T^X \neq T^Y$, more precisely if $T^Y \kappa^{XY} \neq T^X \kappa^{YX}$, then the joint system is out of equilibrium.
	A back-action-free measurement corresponds to $\kappa^{XY} = 0$ but $\kappa^{YX} \neq 0$.

	\begin{figure}
		\includegraphics[width=0.47\textwidth]{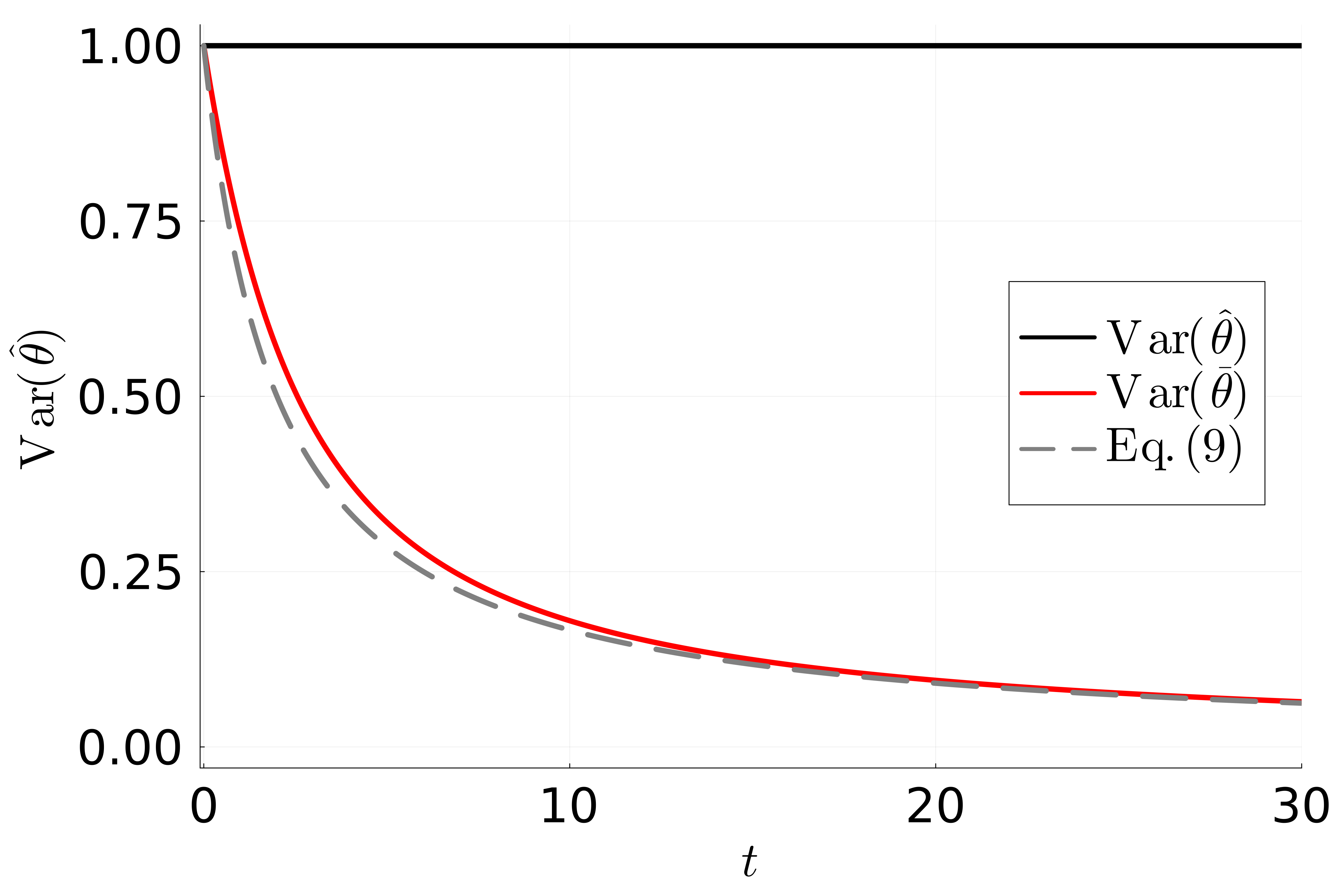}
		\caption{The variance of the time-averaged estimator \eqref{var-time-averaged-estimator} (red) as a function of time. The time-averaged estimator results in a reduced variance compared to the instantaneous estimator $\hat{\theta}$ (black). At the same time, it remains lower bounded by the generalized CRB \eqref{time-average-cramer-rao} (gray dashed), which becomes asymptotically tight at long times.}
		\label{fig-time-average-crb}
	\end{figure}
	
	\paragraph{Time-averaged estimators in linear systems.}
	Remarkably, we find that the generalized CRB \eqref{time-average-cramer-rao} is tight for the corresponding case of \eqref{langevin-linear} to time-averaging.
	For the parameters $T^Y = 0$, $\kappa^{XY} = 0$ and $\kappa_{yx} = -k^Y$, $y(t)$ is the time-integral of $x(t)$,
	\begin{align}
		\dot{y}(t) = r x(t),
	\end{align}
	where $r = k^Y/\gamma^Y$ is a constant of dimension 1/time. 
	The dynamics of $x(t)$, which we assume to be in equilibrium, is unaffected by $y(t)$. 
	Choosing $y(0) = 0$, $\bar{\theta}(y) = y/(t r)$ is then an estimator of $\theta$ since $\av{\bar{\theta}(t)} = \theta$ and equal to the time-average of the estimator $\hat{\theta}(x) = x$.
	Computing the variance of $\bar{\theta}(t)$ from the correlations of $x(t)$ (see Appendix C), we obtain the explicit expression
	\begin{align}\label{var-time-averaged-estimator}
		\text{Var}(\bar{\theta}(t)) = \frac{2 \gamma^X T^X }{(k^X t)^2} \bigg( t - \frac{\gamma^X}{k^X} \Big( 1 - e^{-\frac{k^X}{\gamma^X} t} \Big) \bigg) \simeq \frac{2 \gamma^X T^X}{(k^X)^2 t},
	\end{align}
	where the second expression is the asymptotic long-time behaviour.
	On the other hand, we have $F_\theta^{X,\text{eq}} = k^X/T^X$ and $\mathcal{F}_\theta^{X,\text{eq}} = (k^X)^2/(\gamma^X T^X)$ and therefore from the generalized CRB \eqref{time-average-cramer-rao},
	\begin{align}
		\text{Var}(\bar{\theta}(t)) \geq \frac{1}{\frac{k^X}{T^X} + \frac{(k^X)^2}{2 \gamma^X T^X} t} \simeq \frac{2 \gamma^X T^X}{(k^X)^2 t} .
	\end{align}
	Just like the estimator $\hat{\theta}$ is Cram{\'e}r-Rao-optimal in the sense that $\text{Var}(\hat{\theta}) = 1/F^X_\theta$, the time-averaged estimator $\bar{\theta}$ becomes asymptotically optimal for long times, demonstrating that the bound \eqref{time-average-cramer-rao} can be tight, as illustrated in Fig.~\ref{fig-time-average-crb}.

	\begin{figure}[htbp]
		\centering
		\includegraphics[width=0.48\textwidth]{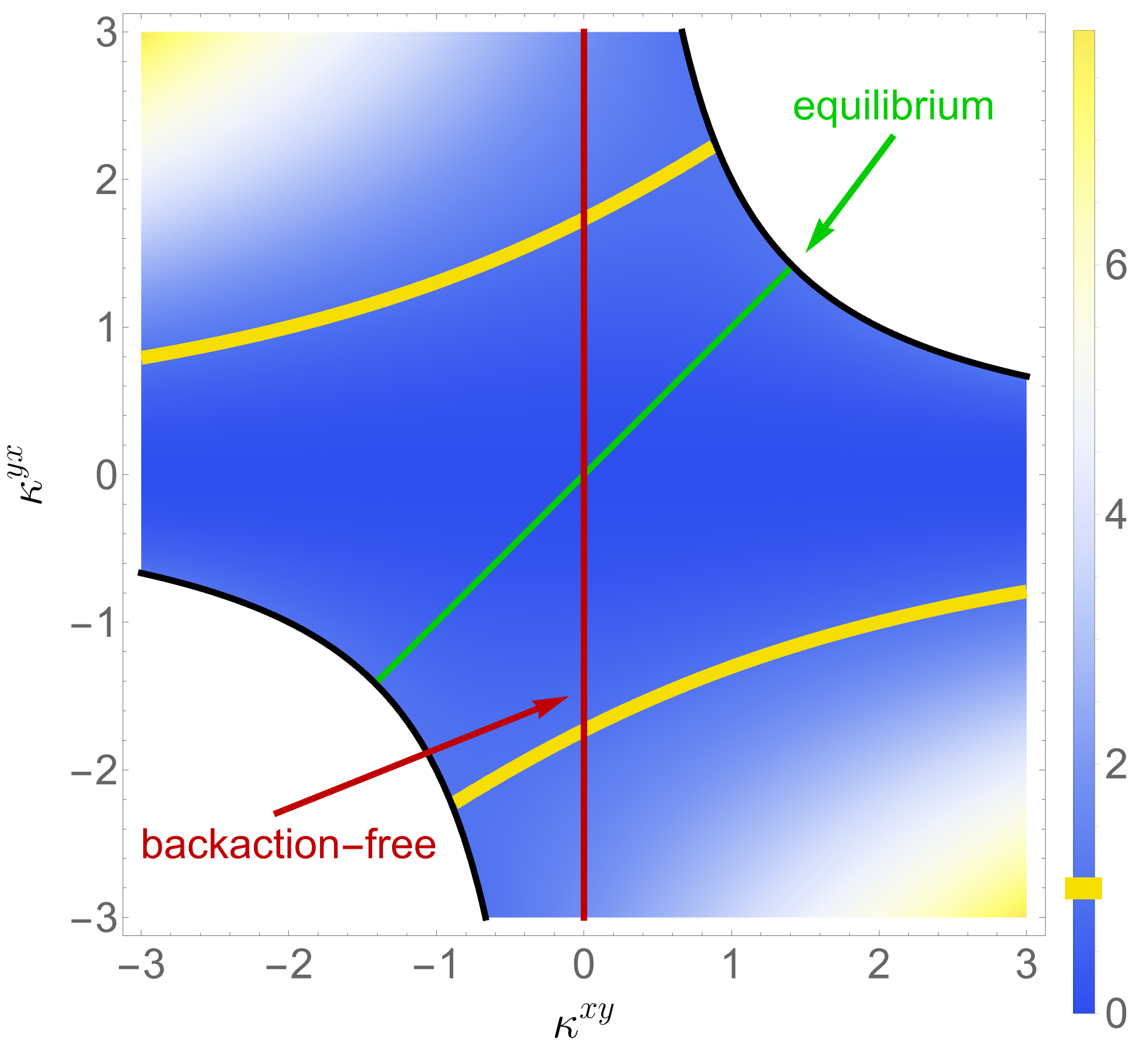}\hfill
		\includegraphics[width=0.48\textwidth]{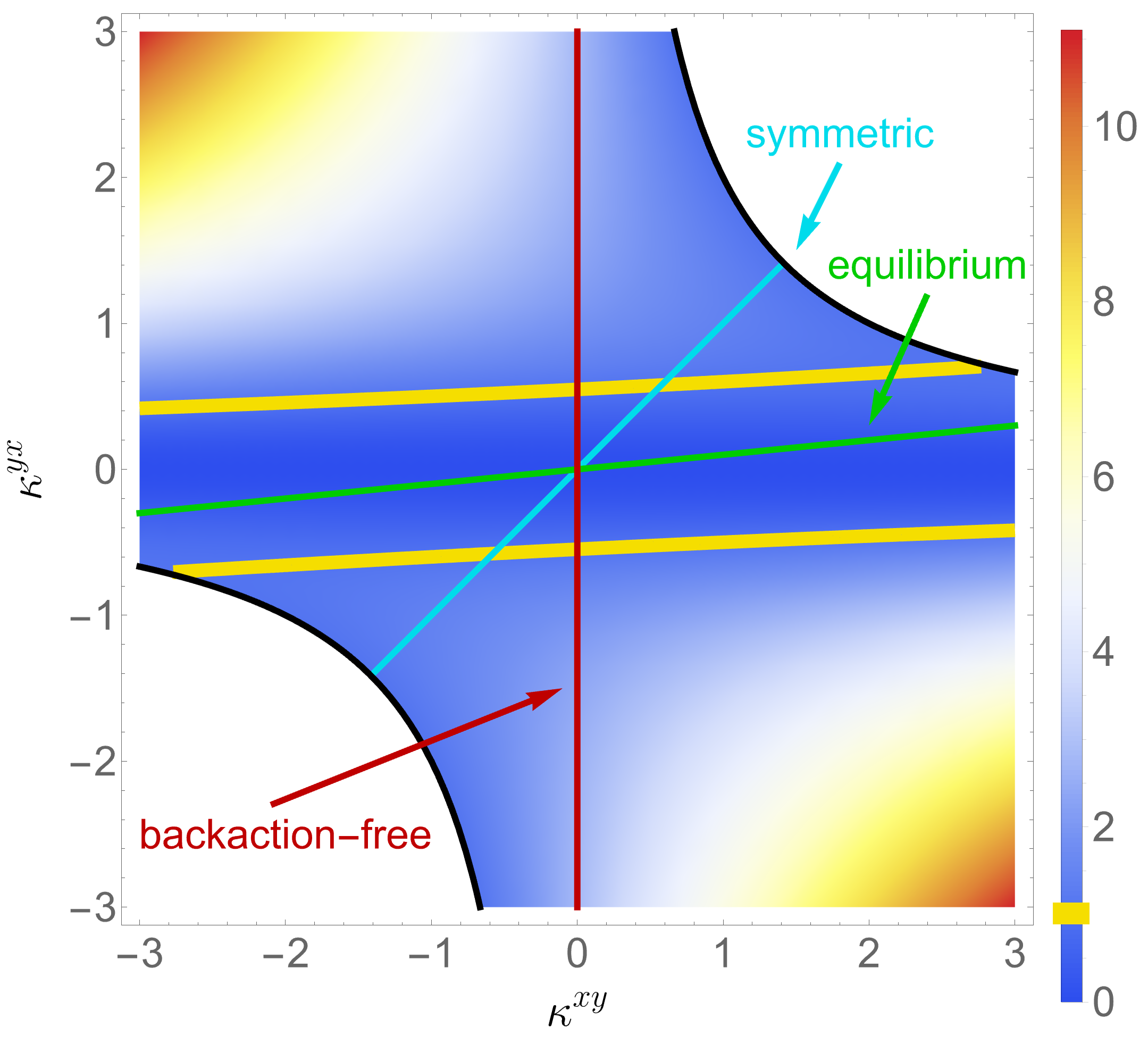}
		\caption{The steady-state ratio between the Fisher information of $Y$ and $X$, $\eta = F^{Y, \mathrm{ss}}_\theta/$ as a function of the coupling constants $\kappa^{XY}$ and $\kappa^{YX}$ for the linear system \eqref{langevin-linear}. 
			The yellow lines indicate the value $1$, beyond these lines, the Fisher information of $Y$ is larger than that of $X$. 
			The green line corresponds to the equilibrium condition $\kappa^{XY} = \kappa^{YX}$, the red line to a back-action-free measurement $\kappa^{XY} = 0$. 
			The black lines indicate the boundary of the stable parameter region where a steady state exists.
			The left panel shows the result for equal temperatures $T^X = T^Y = 1$, while in the right panel, the temperature of the measurement system $Y$ is lower, $T^Y = 0.1$.
			The light blue line in the bottom panel indicates a symmetric coupling $\kappa^{XY} = \kappa^{YX}$.
			The remaining parameter values are $k^X = 2$, $k^Y = 1$ and $\gamma^X = \gamma^Y = 1$.}
		\label{fig-fisher-ratio}
	\end{figure}
	
	\paragraph{Physical measurements: cooling and non-symmetric interactions.}
	From the point of view of \eqref{langevin-linear}, time averaging corresponds to a back-action-free measurement at zero temperature.
	To characterize situations in which $Y$ corresponds to a physical degree of freedom, we introduce the ratio between the Fisher information
	\begin{align}
		\eta = \frac{F^Y_\theta}{F^X_\theta} \label{fisher-ratio} .
	\end{align}
	In equilibrium, we have $\eta \leq 1$, while out of equilibrium $\eta > 1$ can be realized.
	In the following, we focus on the steady-state value $\eta^{\mathrm{ss}}$.
	Due to the linearity of \eqref{langevin-linear}, the steady state (which exists if $k^X > 0$, $k^Y > 0$ and $k^X k^Y > \kappa^{XY} \kappa^{YX}$) is Gaussian with covariance matrix $\bm{\Xi}^\mathrm{ss}$, which can be computed explicitly from the parameters of \eqref{langevin-linear} (see SM \cite{SM}).
	The Fisher information is given by
	\begin{align}
		F^{X, \mathrm{ss}}_\theta = \frac{\big(\partial_\theta \Av{x}_{\theta,\mathrm{ss}}\big)^2}{\left(\bm\Xi^\mathrm{ss}\right)_{xx}},
	\end{align}
	where $\left(\bm\Xi^\mathrm{ss}\right)_{xx}$ denotes the $x$-component of the covariance matrix, and similar for the Fisher information of $Y$.
	
	Fig.~\ref{fig-fisher-ratio}, shows the steady-state ratio $\eta^{\mathrm{ss}}$ as a function of the coupling constants $\kappa^{XY}$ and $\kappa^{YX}$.
	These results suggest two distinct ways of increasing the precision of a measurement of $\theta$ through non-equilibrium conditions.
	Even for symmetric interactions, cooling the measurement system $Y$ reduces its fluctuations \cite{Johnson1928} and therefore allows more accurate estimations than a direct observation of $X$.
	Indeed, this approach is regularly used in practice, where sensors are often cooled to improve their signal-to-noise ratio \cite{Cao2013, Fogliano2021, Han2022}.
	However, cooling is not always feasible, for example in situations where the measured system and measurement system are not spatially well-separated.
	There, enhanced estimation can be realized through non-symmetric coupling between the measured and measurement system, which allows reducing the fluctuations of $Y$ through an effect known as non-reciprocal cooling \cite{Loos2020, Loos2023}.
	Such a coupling, which corresponds to a non-conservative force, leads to a non-equilibrium configuration, allowing for more accurate measurements that permitted by \eqref{fisher-equilibrium} in equilibrium.
	
	It was recently demonstrated that non-symmetric interactions can indeed be realized in experiments, for example via optical interactions \cite{Rieser2022, Reisenbauer2024, Rudolph2024}, feedback \cite{Gavrilov2017} or electro-hydrodynamic interactions \cite{Hara2026}, or in anisotropic plasmas \cite{Lisin2020}.
	In particular, depending on the experimental parameters, the interactions can become completely asymmetric \cite{Rieser2022} such that $f^{XY} = 0$ but $f^{YX} \neq 0$.
	From the point of view of \eqref{langevin}, this means that particle $Y$ is performing a back-action free measurement of the position of particle $X$.
	As shown in Fig.~\ref{fig-fisher-ratio}, the steady-state Fisher information of $Y$ can then exceed the Fisher information of $X$.
	Somewhat counter to intuition, observing the position of particle $Y$ then allows for a more accurate estimation of the position of particle $X$ compared to direct observation.
	
	\paragraph{Discussion.}
	\begin{figure}
		\centering
		\includegraphics[width=0.47\textwidth]{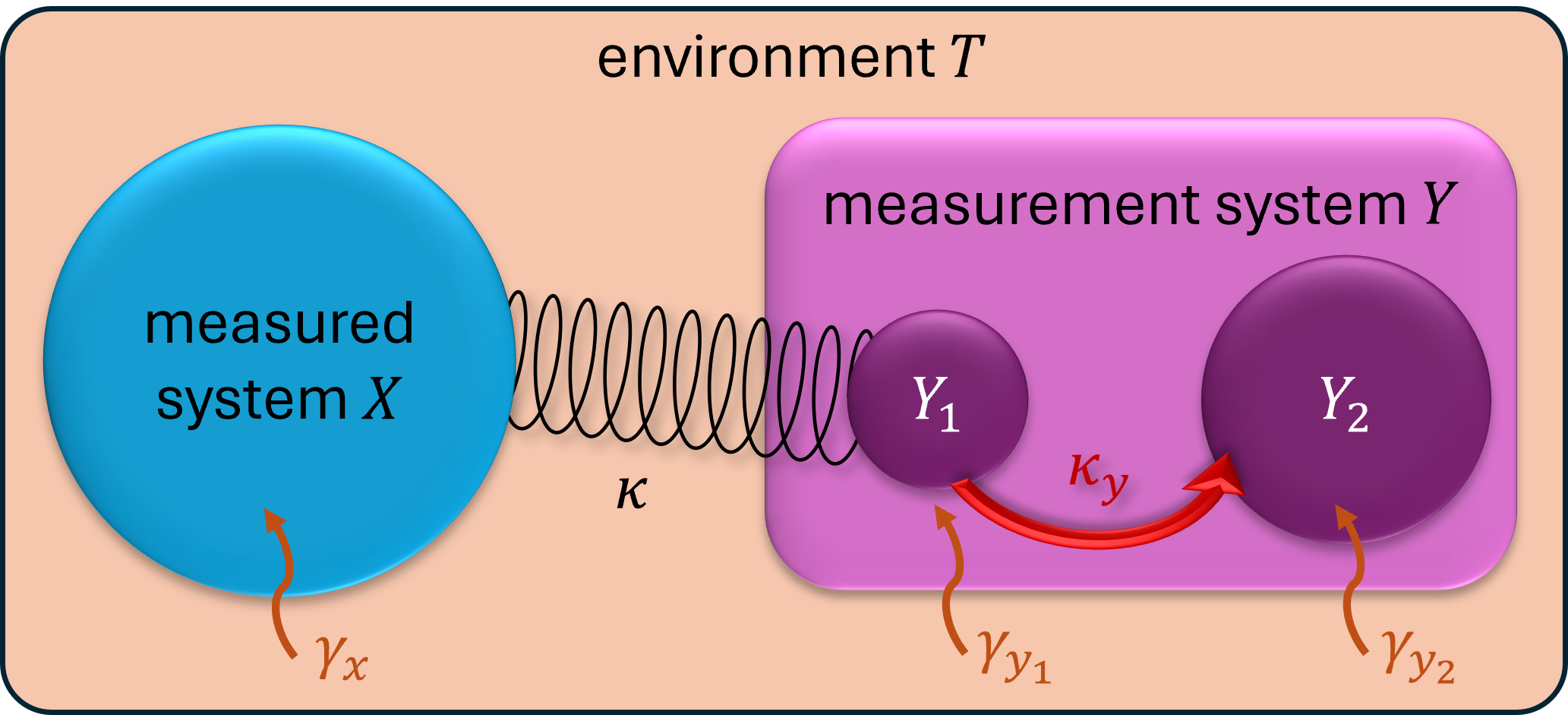}
		\caption{Illustration of the two-component measurement system \eqref{langevin-sensor}.
			The degree of freedom $Y_1$ is coupled to the measured system $X$ via a symmetric coupling with spring constant $\kappa$.
			$Y_2$ then performs a back-actio-free measurement of $Y_2$ through the asymmetric coupling $\kappa^Y$.
			All degrees of freedom are in contact with an environment at temperature $T$ through friction coefficients $\gamma^X$, $\gamma^{Y_1}$ and $\gamma^{Y_2}$, respectively.}
		\label{fig-measurement-setup}
	\end{figure}
	By accumulating information about past states of the measured system, a measurement system may reduce fluctuations of estimators and enable more accurate parameter estimation.
	The corresponding increase in Fisher information is, one the one hand, bounded by the speed limit \eqref{fisher-speed-limit} that depends only on the properties of the measured system and thus limits how much fluctuations can be suppressed by any measurement process, resulting in a generalized CRB for time-averaged estimators \eqref{time-average-cramer-rao}.
	On the other hand, the relation \eqref{fisher-equilibrium} implies that this necessarily requires out-of-equilibrium conditions.
	The latter are also associated with entropy production and, in the context of information thermodynamics, result in information flows between the measured and measurement system \cite{Sagawa2012a, Barato2013, Horowitz2014, Parrondo2015}.
	Whether and how these concepts can be related to the present description in terms of Fisher information is an open question.
	
	In practice, it may be difficult to realize a non-symmetric coupling between the measured and measurement system, in particular when the measured system cannot be controlled directly.
	However, we can combine equilibrium and back-action-free measurements by introducing internal degrees of freedom in the measurement system.
	Concretely, we consider the case where the measurement system has two degrees of freedom $Y_1$ and $Y_2$, as illustrated in Fig.~\ref{fig-measurement-setup}.
	$Y_1$ is coupled to the measured system $X$ via a reciprocal interaction, while $Y_2$ performs a back-action-free measurement of $Y_1$.
	Since $Y_2$ affects neither $X$ nor $Y_1$, these remain in their joint equilibrium state, while the non-equilibrium interactions between $Y_1$ and $Y_2$ allow the Fisher information of $Y$ to exceed that of $X$.
	Remarkably, such a system can be realized even in the linear regime (see Appendix D).
	
	Our results complement existing research demonstrating that non-equilibrium conditions can enhance the signal-to-noise ratio of both biological and artificial sensors \cite{Tu2008, Govern2012, Mehta2012, Lan2012, Dechant2025}.
	While these works focused on the properties of the measurement system $Y$ and its response to given input signals, the present setup makes both the measured and measurement system explicit.
	As a result, the signal, that is, the dependence of $X$ on the parameter $\theta$, is inherently noisy and therefore the measurement precision of $Y$ is fundamentally bounded by the properties of $X$.
	It would be interesting to combine both approaches to obtain more precise informed bounds that take into account the properties of both measured and measurement system.
	
	We anticipate that the present results can be generalized to other classes of dynamics, such as jump processes on a discrete state space or Markovian open quantum systems.
	In the former case, however, notions such as parameter-dependent forces and uncorrelated noises, which are natural for diffusing particle systems, are not straightforward to generalize, making the interpretation of the results more subtle.
	For quantum systems, on the other hand, there is no unique definition of Fisher information and it must be clarified which definition is appropriate.
	More fundamentally, back-action free measurements are prohibited in quantum mechanics and thus the influence of the measurement system on the measured one must be considered carefully.
	
	As a prominent application in quantum systems, the quantum Fisher information is widely used to evaluate the precision of quantum thermometry \cite{staceQuantumLimitsThermometry2010, correaIndividualQuantumProbes2015, depasqualeQuantumThermometry2018, planellaBathInducedCorrelationsEnhance2022}. Along this line, a specific probe–ancilla–bath model was investigated in Ref.~\cite{ullahLowtemperatureQuantumThermometry2023}, reporting an enhanced Fisher information for low temperatures via non-thermal interactions. While this Letter has focused on cases where the parameter of interest appears in the drift term, estimating the reservoir temperature typically involves parameters embedded in the diffusion matrix. The speed limit relation for such a cases is discussed in SM \cite{SM}, however, extending our framework and exploring its implications for quantum metrology, including thermometry remains an important topic for future work.
	
	\paragraph{Acknowledgments.} 
	H.O. was supported by JSPS KAKENHI Grant No. JP26K00022 through a research assistantship.
	A.D. was supported by JSPS KAKENHI (Grants No. 24H00833 and 25K00926).
	This work was supported by JST ERATO, Grant Number JPMJER2302.
	This work was supported by JSPS International Joint Research Program (JRP-LEAD with DFG), Grant Number 20261606.
	
	\bibliography{ref2}
	
	\clearpage
	
	\section{End Matter}
	
	\subsection{Appendix A: Equilibrium measurements}

	Let us assume that $X$ and $Y$ are two interacting systems, jointly characterized by a potential $U_\theta^{xy}$ that can be written as
	\begin{align}
		U_\theta^{xy}(\bm{x},\bm{y}) = U^X_\theta(\bm{x}) + U^Y(\bm{y}) + U^\text{int}(\bm{x},\bm{y}) , \label{hamiltonian}
	\end{align}
	where $U^\text{int}(\bm{x},\bm{y})$ describes the interactions between $X$ and $Y$.
	We assume that only the potential of $X$ depends on the parameter $\theta$.
	The joint equilibrium state of $X$ and $Y$ is then given as the Boltzmann-Gibbs form
	\begin{align}
		p_\theta^{XY,\text{eq}}(\bm{x},\bm{y}) = \frac{\exp\big[-\frac{U_\theta^{xy}(\bm{x},\bm{y})}{T}\big]}{\int d\bm{x} \int d\bm{y} \ \exp\big[-\frac{U_\theta^{xy}(\bm{x},\bm{y})}{T}\big]} \label{boltzmann-gibbs}.
	\end{align}
	
	The Fisher information satisfies the chain rule \cite{Cover2012}
	\begin{align}
		F_\theta^{XY} = F_\theta^{Y \vert X} + F_\theta^X \label{fisher-chain-1},
	\end{align}
	where $F_\theta^{Y \vert X}$ is the Fisher information of the conditional probability density
	\begin{align}
		F_\theta^{Y \vert X} = \Av{\big( \partial_\theta \ln p^{Y \vert X}_\theta \big)^2} .
	\end{align}
	From \eqref{hamiltonian} and \eqref{boltzmann-gibbs}, we obtain
	\begin{align}
		p^{Y \vert X,\text{eq}}_\theta(\bm{y} \vert \bm{x}) = \frac{\exp\big[-\frac{U^Y(\bm{y}) + U^\text{int}(\bm{x},\bm{y})}{T}\big]}{\int d\bm{y} \ \exp\big[-\frac{U^Y(\bm{y}) + U^\text{int}(\bm{x},\bm{y})}{T}\big]} ,
	\end{align}
	which is independent of $\theta$ and, thus, $F_\theta^{Y \vert X,\text{eq}} = 0$ and $F_\theta^{XY,\text{eq}} = F_\theta^{X,\text{eq}}$.
	This means that, in equilibrium, the state of $X$ contains all information about the parameter $\theta$; further measurements of $Y$ do not yield additional information.
	Equivalently, we can decompose the joint Fisher information as $F_\theta^{XY} = F_\theta^{X \vert Y} + F_\theta^Y$.
	Since $F_\theta^{X \vert Y} \geq 0$, this implies $F_\theta^Y \leq F_\theta^{XY}$ and, thus, $F_\theta^{Y,\text{eq}} \leq F_\theta^{X,\text{eq}}$ in equilibrium, which is \eqref{fisher-equilibrium}.
	
	\subsection{Appendix B: Derivation of \eqref{fisher-speed-limit}}
	We start by calculating the time-derivative of the joint Fisher information $d_t F^{XY}_\theta(t) = d_t \av{(\partial_\theta \ln p^{XY}_\theta )^2}$.
	We replace the time-derivative of the joint probability density using the Fokker-Planck equation
	\begin{align}
		\partial_t &p_\theta^{XY} = - \grad_z \cdot \big( p_\theta^{XY} \big( \bm{a}_\theta - \bm{B} \grad_z \ln p_\theta^{XY} \big) \Big),
	\end{align}
	where we denote by $\bm{z} = (\bm{x},\bm{y})$ the joint configuration and by $\grad_z = (\grad_x,\grad_y)$ the corresponding gradient operator, and defined the drift vector and diffusion matrix as
	\begin{align}
		\bm{a}_\theta(\bm{x},\bm{y}) &= \begin{pmatrix} \bm{\gamma}^{x,-1} \big( \bm{f}^X_\theta(\bm{x}) + \bm{f}^{XY}(\bm{x},\bm{y}) \big) \\ \bm{\gamma}^{y,-1} \big( \bm{f}^Y(\bm{y}) + \bm{f}^{YX}(\bm{x},\bm{y}) \big) \end{pmatrix} \\
		\bm{B} &= \begin{pmatrix} \bm{\gamma}^{x,-1} T^X & \bm{0} \\ \bm{0} & \bm{\gamma}^{y,-1} T^Y \end{pmatrix}. \n
	\end{align}
	Using this, the time-derivative of the joint Fisher information can be calculated as
	\begin{align}
		d_t F^{XY}_\theta(t) &= -2 \Av{\partial_\theta \grad_z \ln p^{XY}_\theta \cdot \bm{B} \partial_\theta \grad_z \ln p^{XY}_\theta} \nn
		& + 2 \Av{\partial_\theta \grad_z \ln p^{XY}_\theta \cdot \partial_\theta \bm{a}_\theta} .
	\end{align}
	The calculation closely follows Appendix D of Ref.~\cite{Ito2020}; for convenience it is provided in detail in SM \cite{SM}.
	We now use that $\bm{a}_{\theta}(\bm{x},\bm{y})$ only depends on $\theta$ via $\bm{f}^X_\theta(\bm{x})$ and that the diffusion matrix is block-diagonal to write this as
	\begin{align}
		d_t F^{XY}_\theta(t) &= - 2 \Av{\partial_\theta \grad_y \ln p^{XY}_\theta \cdot \bm{\gamma}^{y,-1} T^Y \partial_\theta \grad_y \ln p^{XY}_\theta} \nn
		& \quad -2 \Av{\partial_\theta \grad_x \ln p^{XY}_\theta \cdot \bm{\gamma}^{x,-1} T^X \partial_\theta \grad_x \ln p^{XY}_\theta} \nn
		& \quad + 2 \Av{\partial_\theta \grad_x \ln p^{XY}_\theta \cdot \bm{\gamma}^{x,-1} \partial_\theta \bm{f}^X_\theta} \label{fisher-derivative} .
	\end{align}
	We can bound the magnitude of the third term by using the Cauchy-Schwarz inequality,
	\begin{align}
		&\Av{\partial_\theta \grad_x \ln p^{XY}_\theta \cdot \bm{\gamma}^{x,-1} \partial_\theta \bm{f}_\theta}^2 \leq \mathcal{B}_\theta^X(t) \mathcal{F}_\theta^X(t),
	\end{align}
	where we defined
	\begin{subequations}
		\begin{align}
			\mathcal{B}_\theta^X(t) &= \Av{\partial_\theta \grad_x \ln p^{XY}_\theta \cdot \bm{\gamma}^{x,-1} T^X \partial_\theta \grad_x \ln p^{XY}_\theta}, \\
			\mathcal{F}_\theta^X(t) &= \frac{1}{T^X} \Av{\partial_\theta \bm{f}_\theta \cdot \bm{\gamma}^{x,-1} \partial_\theta \bm{f}_\theta}. \label{fisher-rate-definition-2} 
		\end{align}
	\end{subequations}
	Since the term in the first line of \eqref{fisher-derivative} is negative, we obtain the upper bound on the rate of change of the Fisher information
	\begin{align}
		d_t F^{XY}_\theta(t) &\leq - 2 \mathcal{B}_\theta^X(t) + 2 \sqrt{\mathcal{B}_\theta^X(t) \mathcal{F}_\theta^X(t)} .
	\end{align}
	Maximizing the right-hand side with respect to $\mathcal{B}_\theta^X(t)$, we obtain
	\begin{align}
		d_t F^{XY}_\theta(t) \leq \frac{1}{2} \mathcal{F}_\theta^X(t).
	\end{align}
	We also note that, even when the measurement is not back-action free and $X$ is not in equilibrium, the Fisher information of $Y$ always obeys the more general speed limit,
	\begin{align}
		F_\theta^Y(t) \leq F_\theta^X(0) + \int_0^\tau dt' \ \frac{1}{2} \mathcal{F}_\theta^X(t'),
	\end{align}
	where $\mathcal{F}_\theta^X(t)$ is given by \eqref{fisher-rate-definition-2}, which generally depends on time due to the back-action of $Y$ on $X$, and, therefore, implicitly on the properties of the measurement system $Y$.
	As in the derivation of \eqref{fisher-speed-limit}, we assumed that $Y$ initially carries no information about $\theta$, that is, $F_\theta^Y(0) = 0$ and $F_\theta^{XY}(0) = F_\theta^X(0)$.

	\subsection{Appendix C: Derivation of \eqref{var-time-averaged-estimator}}
	
	For the case of time-averaged estimators, our general setup for two-dimensional linear systems \eqref{langevin-linear} reduces to
	\begin{align}
		\gamma^X \dot{x}(t) &=  -k^X (x(t) - \theta) + \sqrt{2 \gamma^X T^X} \xi^X(t), \nn
		\dot{y}(t) &= r x(t),
	\end{align}
	where $r = k^Y/\gamma^Y$.
	This is formally solved by 
	\begin{align}
		x(t) &= x(0)e^{-k^Xt/\gamma^X}\nn
		&+ \int_{0}^{t} dt' \, e^{-k^X(t-t')/\gamma^X}\left(\frac{k^X\theta}{\gamma^X} + \sqrt{\frac{2T}{\gamma^X}} \xi^X(t') \right), \label{formal-x} \\
		y(t) &= y(0) + \int_0^t dt' \ x(t'). \n
	\end{align}
	Averaging and using $\langle x(0)\rangle = \theta$ and $\langle y(0)\rangle = 0$ as in the main text, we obtain
	\begin{align}
		\langle x(t)\rangle &= \theta, \qquad \langle y(t)\rangle = r\theta t.
	\end{align}
	We then obtain the two-time correlation of $x(t)$ as
	\begin{align}
		\langle &x(s)x(s')\rangle = \theta^2 + \frac{T}{k^X}e^{-k^X(s-s')/\gamma^X},
	\end{align}
	where we substituted the initial condition $\langle x(0)\rangle = \theta$, $\langle x(0)^2 \rangle = \theta^2 + T/k^X$ and assumed $s\geq s'$ without loss of generality. 
	The variance of $y(t)$ is computed as
	\begin{align}
		\text{Var}&(y(t)) = 2r^2\int_0^tds\int_0^{s}ds'\, \langle x(s)x(s')\rangle - \langle y(t) \rangle^2 \nn
		&= \frac{2r^2\gamma^X Tt}{(k^X)^2} + \frac{2r^2(\gamma^X)^2Te^{-\frac{k^Xt}{\gamma^X}}}{(k^X)^3}-\frac{2 r^2(\gamma^X)^2 T}{(k^X)^3}
	\end{align}
	which immediately leads to \eqref{var-time-averaged-estimator}.
	
	\section{Appendix D: Disturbance-free non-equilibrium sensing}
	
	We consider the case where the measurement system has two degrees of freedom $Y_1$ and $Y_2$ (Fig.~\ref{fig-measurement-setup}).
	$Y_1$ is coupled to the measured system $X$ via a symmetric coupling potential $U^\text{int} = \kappa (x-y_1)^2/2$.
	This coupling does not affect the equilibrium distribution of $X$, $p_\theta^{X,\text{eq}}(x) \propto e^{-U_\theta^X(x)/T}$.
	We further introduce a non-reciprocal interaction, so that $Y_2$ performs a back-action-free measurement of $Y_1$.
	
	In the context of the linear system \eqref{langevin-linear}, we write 
	\begin{align}
		\gamma^X \dot{x} &=  -k^X (x - \theta) - \kappa (x-y_1) + \sqrt{2 \gamma^X T} \xi^X, \nn
		\gamma^{Y_1} \dot{y}_1 &= -\kappa (y_1 - x)  + \sqrt{2 \gamma^{Y_1} T} \xi^{y_1} , \label{langevin-sensor} \\
		\gamma^{Y_2} \dot{y}_2 &= -k^Y y_2 + \kappa^Y y_1 + \sqrt{2 \gamma^{Y_2} T} \xi^{y_2} \n ,%
	\end{align} 
	where $\kappa$ is the symmetric coupling between $X$ and $Y_1$, while $\kappa^Y$ is strength of the back-action-free measurement.
	Here, we assume that all parts of the system are in contact with the same thermal reservoir, while the friction coefficients may be different.
	If $k^X, k^Y, \kappa > 0$, then this system has a stable steady state, whose probability density can be computed explicitly (see SM \cite{SM}).

	The information that $Y$ carries about $\theta$ is maximized when both $\kappa$ and $\kappa^Y$ are large.
	In this strong-coupling limit, we obtain for the ratio \eqref{fisher-ratio} in the steady state
	\begin{align}
		\eta^{\mathrm{ss}} \simeq 1 + \frac{k^X \gamma^{Y_2}}{k^Y(\gamma^X + \gamma^{Y_1})} + O(1/\kappa,1/\kappa^Y).
	\end{align}
	Thus, the Fisher information of $Y$ becomes large when either $k^Y$ is small or $\gamma^{Y_2}$ is large.
	Since the former corresponds to a freely diffusing $Y_2$ without a steady state, we focus on the latter.
	A large Fisher information then requires $\gamma^{Y_2} \gg \gamma^{Y_1}$, which can be realized, e.g., by particles of different sizes, since friction scales with particle size.
	Physically, $Y$ is a good sensor if the degree of freedom $Y_1$ coupled to the system is fast, while the slow $Y_2$ performs a back-action-free measurement of $Y_1$.
	
\begin{figure}
		\centering
		\includegraphics[width=0.47\textwidth]{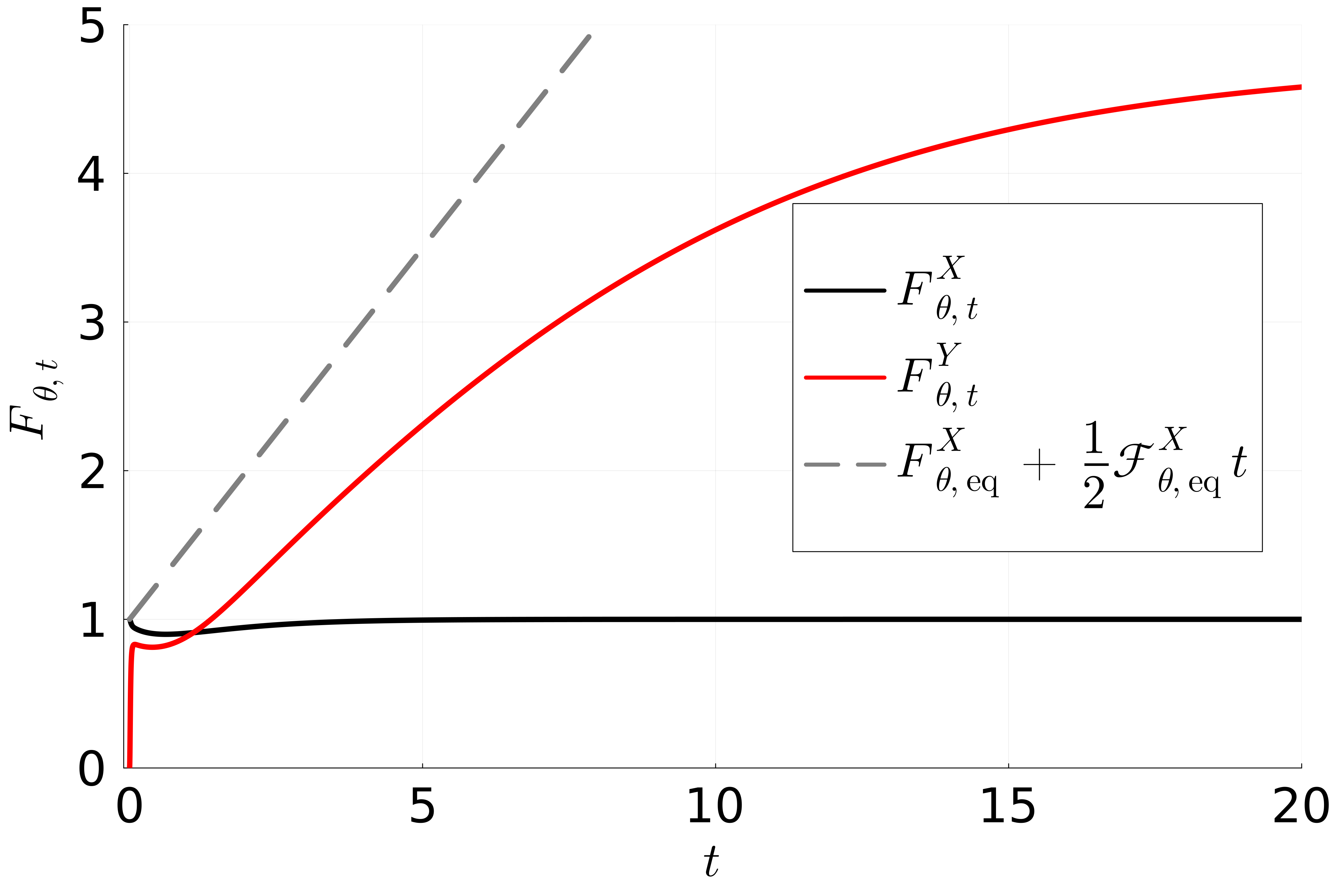}
		\caption{Fisher information of $X$ (black) and $Y$ (red) versus time for the sensor \eqref{langevin-sensor}. Owing to its non-symmetric interactions, $Y$ eventually surpasses the equilibrium value of $X$ ($=1$), but always remains below the speed limit \eqref{fisher-speed-limit} (dashed). Parameters: $k^X = k^Y = 1$, $\kappa = \kappa^Y = 10$, $\gamma^X = 1$, $\gamma^{Y_1} = 0.2$, $\gamma^{Y_2} = 5$, $T = 1$.}
		\label{fig-measurement}
	\end{figure}

	Although $Y$ allows an arbitrarily precise measurement of $\theta$ for $\gamma^{Y_2} \gg \gamma^X + \gamma^{Y_1}$, the growth of $F_\theta^Y(t)$ is bounded by the speed limit \eqref{fisher-speed-limit} (Fig.~\ref{fig-measurement}).
	This implies a trade-off: a large $\gamma^{Y_2}$ increases the steady-state information but slows relaxation, so information cannot be acquired faster than \eqref{fisher-speed-limit} allows.
	
	\clearpage
	
	\onecolumngrid
	
	\newcommand{\eq}{\text{eq}}
	\newcommand{\stst}{\text{ss}}
	\newcommand{\const}{(\text{const.})}
	
	\renewcommand{\theequation}{S\arabic{equation}}
	\renewcommand{\thefigure}{S\arabic{figure}}
	\renewcommand{\thetable}{S\arabic{table}}
	\renewcommand{\thesection}{S\Roman{section}}
	
	\setcounter{secnumdepth}{2}
	\setcounter{equation}{0}
	
	\section*{Supplemental Material}
	
	\section{Proof of Eq.~(23)}
	
	Here we derive the expression for the time-derivative of Fisher information.
	We first consider a general Langevin equation with parameter dependent drift vector and diffusion matrix and then specialize to the case where we subdivide the system into a measured system $X$ and a measurement system $Y$.
	We start from the overdamped Langevin equation
	\begin{align}
		\dot{\bm{z}}(t) = \bm{a}_\theta(\bm{z}(t)) + \sqrt{2 \bm{B}_\theta} \bm{\xi}(t), \label{langevin-sup}
	\end{align}
	where $\bm{a}_\theta(\bm{z})$ is drift vector and $\bm{B}_\theta$ the positive definite diffusion matrix, both of which can depend on a parameter $\theta$.
	The corresponding Fokker-Planck equation for the probability density $p_\theta(\bm{z}, t)$ is
	\begin{align}
		\partial_t p_\theta(\bm{z},t) = - \grad_z \cdot \Big( p_\theta(\bm{z}, t) \big( \bm{a}_\theta(\bm{z}, t) &- \bm{B}_\theta \grad_z \ln p_\theta(\bm{z}, t) \big) \Big) = - \grad_z \cdot \Big( \bm{\nu}_\theta(\bm{z}, t) p_\theta(\bm{z}, t) \Big) \nn
		\text{with} \quad \bm{\nu}_\theta(\bm{z}, t) = \bm{a}_\theta(\bm{z}, t) &- \bm{B}_\theta \grad_z \ln p_\theta(\bm{z}, t) . \label{fpe-sup}
	\end{align}
	We now compute the time-derivative of the Fisher information with respect to $\theta$
	\begin{align}
		d_t F_\theta(t) = d_t \int d\bm{z} \ \big( \partial_\theta \ln p_\theta(\bm{z}, t) \big)^2 p_\theta(\bm{z}, t) .
	\end{align}
	We use that
	\begin{align}
		\partial_t \partial_\theta \ln p_\theta(\bm{z}, t) = \frac{\partial_t \partial_\theta p_\theta(\bm{z}, t)}{p_\theta(\bm{z}, t)} - \partial_t  \ln p_\theta(\bm{z}, t) \partial_\theta \ln p_\theta(\bm{z}, t)
	\end{align}
	and
	\begin{align}
		\partial_t \partial_\theta p_\theta(\bm{z}, t) = - \grad_z \cdot \big( p_\theta(\bm{z}, t) \partial_\theta \bm{\nu}_\theta(\bm{z}, t) + \bm{\nu}_\theta(\bm{z}, t) \partial_\theta p_\theta(\bm{z}, t) \big) .
	\end{align}
	With this, we obtain
	\begin{align}
		d_t F_\theta(t) &= \int d\bm{z} \ \bigg( 2 \partial_\theta \ln p_\theta(\bm{z}, t) \Big( - \grad_z \cdot \big( p_\theta(\bm{z}, t) \partial_\theta \bm{\nu}_\theta(\bm{z}, t) + \bm{\nu}_\theta(\bm{z}, t) \partial_\theta p_\theta(\bm{z}, t) \big) - \partial_t  \ln p_\theta(\bm{z}, t) \partial_\theta \ln p_\theta(\bm{z}, t) p_\theta(\bm{z}, t) \Big) \\
		& \hspace{3cm} + \big( \partial_\theta \ln p_\theta(\bm{z}, t) \big)^2 \partial_t p_\theta(\bm{z}, t) \bigg). \n
	\end{align}
	We integrate by parts with respect to $\bm{z}$ in the first term to get
	\begin{align}
		d_t F_\theta(t) &= \int d\bm{z} \ \bigg( 2 \grad_z \partial_\theta \ln p_\theta(\bm{z}, t) \cdot \big( p_\theta(\bm{z}, t) \partial_\theta \bm{\nu}_\theta(\bm{z}, t) + \bm{\nu}_\theta(\bm{z}, t) \partial_\theta p_\theta(\bm{z}, t) \big) - \big( \partial_\theta \ln p_\theta(\bm{z}, t) \big)^2 \partial_t p_\theta(\bm{z}, t) \bigg) \nn
		&= 2 \int d\bm{z} \ \bigg( \grad_z \partial_\theta \ln p_\theta(\bm{z}, t) \cdot \partial_\theta \bm{\nu}_\theta(\bm{z}, t) + \partial_\theta \ln p_\theta(\bm{z}, t) \grad_z \partial_\theta \ln p_\theta(\bm{z}, t) \cdot \bm{\nu}_\theta(\bm{z}, t) \\
		&\hspace{3cm} -  \partial_\theta \ln p_\theta(\bm{z}, t) \partial_\theta \grad_z \ln p_\theta(\bm{z}, t)  \cdot \bm{\nu}_\theta(\bm{z}, t)  \bigg) p_\theta(\bm{z}, t) \n ,
	\end{align}
	where we used \eqref{fpe-sup} and integrated by parts with respect to $\bm{z}$ in the last term.
	The second and third term cancel, so we obtain
	\begin{align}
		d_t F_\theta(t) &= 2 \int d\bm{z} \ \big( \grad_z \partial_\theta \ln p_\theta(\bm{z}, t) \cdot \partial_\theta \bm{\nu}_\theta(\bm{z}, t) \big) p_\theta(\bm{z}, t) .
	\end{align}
	Plugging in the definition of the local mean velocity $\bm{\nu}_\theta(\bm{z}, t)$, we can write this as
	\begin{align}
		d_t F_\theta(t) &= 2 \int d\bm{z} \ \Big( \grad_z \partial_\theta \ln p_\theta(\bm{z}, t) \cdot \big(\partial_\theta \bm{a}_\theta(\bm{z}) - \partial_\theta \bm{B}_\theta \grad_z \ln p_\theta(\bm{z}, t) \big) \Big) p_\theta(\bm{z}, t) \\
		&\hspace{3cm} - 2 \int d\bm{z} \ \Big( \grad_z \partial_\theta \ln p_\theta(\bm{z}, t) \cdot \bm{B}_\theta \grad_z \partial_\theta \ln p_\theta(\bm{z}, t) \Big) p_\theta(\bm{z}, t) . \n
	\end{align}
	The term in the first line is proportional to the derivative of the drift coefficient and diffusion matrix with respect to $\theta$, while the second term is negative due to the fact that the diffusion matrix is positive definite.
	Intuitively, the negative second term quantifies the loss of information about $\theta$ due to the effect of noise.
	By contrast, the first term can be positive or negative and quantifies how the information about $\theta$ changes due to the explicit parameter-dependence of the dynamics.
	If the diffusion matrix does not depend on $\theta$, then this reduces to Eq.~(23) in Appendix B.
	
	As in the main text, we now decompose the system into a measured system $X$ and measurement system $Y$, $\bm{z} = (\bm{x},\bm{y})$.
	We assume that only the dynamics of $X$ explicitly depend on the parameter $\theta$ and that the noises acting on $X$ and $Y$ are independent.
	We then have
	\begin{align}
		d_t F^{XY}_\theta(t) &= 2 \int d\bm{z} \ \Big( \grad_x \partial_\theta \ln p^{XY}_\theta(\bm{z}, t) \cdot \big(\partial_\theta \bm{a}^{X}_\theta(\bm{z}) - \partial_\theta \bm{B}^{X}_\theta \grad_x \ln p^{XY}_\theta(\bm{z}, t) \big) \Big) p^{X,Y}_\theta(\bm{z}, t) \\
		& - 2 \int d\bm{z} \ \Big( \grad_x \partial_\theta \ln p^{XY}_\theta(\bm{z}, t) \cdot \bm{B}^{X}_\theta \grad_z \partial_\theta \ln p^{XY}_\theta(\bm{z}, t) + \grad_y \partial_\theta \ln p^{XY}_\theta(\bm{z}, t) \cdot \bm{B}^{Y}_\theta \grad_y \partial_\theta \ln p^{XY}_\theta(\bm{z}, t) \Big) p^{XY}_\theta(\bm{z}, t) . \n
	\end{align}
	Thus, while both $X$ and $Y$ contribute to the loss of information due to noise, the Fisher information can increase solely due to the $\theta$-dependence of the drift vector and diffusion matrix of $X$.
	Since both terms in the second line are negative, we further have
	\begin{align}
		d_t F^{XY}_\theta(t) &\leq 2 \int d\bm{z} \ \Big( \grad_x \partial_\theta \ln p^{XY}_\theta(\bm{z}, t) \cdot \big(\partial_\theta \bm{a}^{X}_\theta(\bm{z}) - \partial_\theta \bm{B}^{X}_\theta \grad_x \ln p^{XY}_\theta(\bm{z}, t) \big) \Big) p^{X,Y}\theta(\bm{z}, t) \\
		&\hspace{2cm} - 2 \int d\bm{z} \ \Big( \grad_x \partial_\theta \ln p^{XY}_\theta(\bm{z}, t) \cdot \bm{B}^{X}_\theta \grad_z \partial_\theta \ln p^{XY}_\theta(\bm{z}, t) \Big) p^{XY}_\theta(\bm{z}, t) . \nn
		&\leq \frac{1}{2} \int d\bm{z} \ \big( \partial_\theta \bm{a}^{X}_\theta(\bm{z}) - \partial_\theta \bm{B}^{X}_\theta \grad_x \ln p^{XY}_\theta(\bm{z}, t) \big) \cdot \big(\bm{B}^{X}_\theta \big)^{-1} \big( \partial_\theta \bm{a}^{X}_\theta(\bm{z}) - \partial_\theta \bm{B}^{X}_\theta \grad_x \ln p^{XY}_\theta(\bm{z}, t) \big) \n .
	\end{align}
	If the diffusion matrix $B^{X}$ is independent of $\theta$, this reduces to Eq.~(28) in Appendix B.
	
	\section{Linear Langevin systems and Fisher information}
	\subsection{Lyapunov equation} \label{sec-lyapunov}
	
	In this subsection, we describe the solution of linear Langevin equations. We consider an overdamped Langevin system with a negative-definite drift matrix $\bm A$, given by
	\begin{equation}\label{lin-overdamp}
		\dot{\bm x}(t) = \bm A\bm x(t) + \sqrt{2\bm D}\bm\xi(t),
	\end{equation}
	where $\bm D$ is the positive-semidefinite diffusion matrix which is time-independent and $\bm\xi(t)$ is Gaussian white noise satisfying $\av{\xi_i(t)}=0$ and $\av{\xi_i(t)\xi_j(t')} = \delta_{ij}\delta(t-t')$. The system \eqref{lin-overdamp} is formally solved as
	\begin{equation}\label{formal-sol}
		\bm x(t) = e^{\bm A t}\bm x(0) + \int_{0}^{t} dt' \, e^{\bm A(t-t')}\sqrt{2\bm D}\bm\xi(t').
	\end{equation}
	The ensemble average of $\bm x$ is simply given by
	\begin{equation}
		\av{\bm x(t)} = e^{\bm A t}\av{\bm x(0)},
	\end{equation}
	which converges to $\bm 0$ as $t \to \infty$; $\av{\bm x}_{\rm ss} = \bm 0$. From \eqref{formal-sol}, we obtain the two-time correlation for $t\geq s$ as
	\begin{equation}
		\av{\bm x(t)\bm x(s)^\top} = e^{\bm A t}\av{\bm x(0)\bm x(0)^\top}e^{\bm A^\top s} + \int_{0}^{s}dt'\, e^{\bm A(t - t')}\bm B e^{\bm A^\top(s-t')},
	\end{equation}
	where we have defined $\bm B = \sqrt{2\bm D}\sqrt{2 \bm D^\top}$. We then obtain the steady-state covariance matrix:
	\begin{equation}\label{sigma-ss}
		\bm{\Xi}^{\rm ss} = \av{\bm x(t)\bm x(t)^\top}_{\rm ss} = \int_{-\infty}^{t}dt'\, e^{\bm A(t - t')}\bm B e^{\bm A^\top(t-t')}.
	\end{equation} 
	From \eqref{sigma-ss}, we can derive the following relation:
	\begin{align}
		\bm{A}\bm{\Xi}^{\rm ss} + \bm{\Xi}^{\rm ss}\bm{A}^\top 
		&= \int_{-\infty}^{t} dt' \left[ \bm{A} e^{\bm{A}(t-t')} \bm{B} e^{\bm{A}^\top(t-t')} + e^{\bm{A}(t-t')} \bm{B} e^{\bm{A}^\top(t-t')} \bm{A}^\top \right] \notag \\
		&= -\int_{-\infty}^{t} dt' \, \frac{d}{dt'} \left[ e^{\bm{A}(t-t')} \bm{B} e^{\bm{A}^\top(t-t')} \right] \notag \\
		&= -\left[ e^{\bm{A}(t-t')} \bm{B} e^{\bm{A}^\top(t-t')} \right]_{t'=-\infty}^{t' = t} \notag \\
		&= -\bm{B}.
	\end{align}
	Here, we have used the fact that $\lim_{t' \to -\infty} e^{\bm{A}(t-t')} = \bm{0}$ because all eigenvalues of $\bm{A}$ have strictly negative real parts. Consequently, the steady-state covariance matrix satisfies
	\begin{equation}\label{lyapunov-ss}
		\bm{A}\bm{\Xi}^{\rm ss} + \bm{\Xi}^{\rm ss}\bm{A}^\top + \bm{B} = \bm{0}.
	\end{equation}
	This algebraic relation is known as the continuous-time Lyapunov equation \cite{Risken1986, Gardiner2002}, where $\bm{\Xi}^{\rm ss}$ is given as its solution. The time evolution of the covariance matrix is given by
	\begin{equation}
		\bm{\Xi}(t) = \av{\bm{x}(t)\bm{x}(t)^\top} = e^{\bm A t}\av{\bm{x}(0)\bm{x}(0)^\top}e^{\bm A^\top t} + \int_{0}^{t}dt'\, e^{\bm A(t - t')}\bm B e^{\bm A^\top(t-t')}.
	\end{equation}
	Differentiating this equation with respect to time yields
	\begin{equation}\label{lyapunov-time-gen}
		\dot{\bm{\Xi}}(t) = \bm A\bm{\Xi}(t) + \bm{\Xi}(t)\bm A^\top + \bm B.
	\end{equation}
	\eqref{lyapunov-time-gen} implies that in linear systems, the time evolution of the second moments is expressed in a closed form. This means that if the initial state is Gaussian, the state remains Gaussian for all $t$.
	
	\subsection{Fisher information in Gaussian states}\label{sec-lin-fisher}
	Next, we derive explicit formulations of the Fisher information in linear Langevin systems. As in the main text, we assume that only the drift term depends on the parameter under consideration $\theta$:
	\begin{equation}
		\dot{\bm x}(t) = \bm A_\theta \bm x(t) + \bm b_\theta + \sqrt{2\bm D}\bm\xi(t),
	\end{equation}
	where $\bm A_\theta$ is negative-definite. The steady-state solution for $\av{\bm x(t)}$ is obtained as
	\begin{equation}
		\av{\bm x}_{\theta, \rm ss} = - \bm{A}_\theta^{-1}\bm b_\theta,
	\end{equation}
	which explicitly depends on $\theta$. The time evolution of the ensemble-averaged dynamics $\av{\bm x(t)}_\theta$ is given by
	\begin{equation}
		\av{\bm x(t)}_\theta = \av{\bm x}_{\theta, \rm ss} + e^{\bm A_\theta t}(\av{\bm x(0)}-\av{\bm x}_{\theta, \rm ss}).
	\end{equation}
	As described in the previous subsection, if the initial state is Gaussian, the state remains Gaussian for all $t$. The time evolution of the covariance matrix is governed by the Lyapunov equation (see \eqref{lyapunov-time-gen}):
	\begin{equation}
		\dot{\bm{\Xi}}_{\theta}(t) = \bm A_\theta \bm{\Xi}_\theta(t) + \bm{\Xi}_\theta(t) \bm A_\theta^\top + \bm B. 
	\end{equation}
	Therefore, the probability distribution at time $t$ is expressed as
	\begin{equation}
		p_\theta(\bm x, t) \propto \exp\left(-\frac{1}{2}(\bm x-\av{\bm x(t)}_\theta)^\top\bm{\Xi}_\theta^{-1}(t)(\bm x-\av{\bm x(t)}_\theta)\right).
	\end{equation}
	Thus, the Fisher information with respect to $\theta$ is obtained as
	\begin{equation}\label{FI-lin-gen}
		\begin{aligned}
			F_\theta (t) &= \Av{(\partial_\theta\ln p_\theta(\bm x, t))^2} = -\Av{\partial^2_\theta\ln p_\theta(\bm x, t)} \\
			&= \frac{1}{2}\operatorname{Tr}\left[ \bm{\Xi}_\theta^{-1}(t) (\partial_\theta \bm{\Xi}_\theta(t)) \bm{\Xi}_\theta^{-1}(t) (\partial_\theta \bm{\Xi}_\theta(t)) \right] + (\partial_\theta\av{\bm x(t)}_\theta)^\top \bm{\Xi}_\theta^{-1}(t) (\partial_\theta\av{\bm x(t)}_\theta).
		\end{aligned}
	\end{equation}
	When $\bm A$ is independent of $\theta$ (implying $\bm{\Xi}(t)$ is independent of $\theta$) as considered in the main text, \eqref{FI-lin-gen} reduces to
	\begin{equation}
		F_\theta(t) = (\partial_\theta\av{\bm x(t)}_\theta)^\top \bm{\Xi}^{-1}(t) (\partial_\theta\av{\bm x(t)}_\theta).
	\end{equation}
	
	\section{Condition for $F_\theta^{X, \rm ss}\leq F_\theta^{Y, \rm ss}$ in two-dimensional linear systems}
	
	In the main text, we have focused mainly on the equilibrium and back-action-free measurements, here we consider more general cases; the back-action of measurements may not be negligible. Thus, the system is given by the two-dimensional linear Langevin equation:
	\begin{equation}\label{2d-lin-model-2}
		\begin{aligned}
			\dot{x}(t) &= -\mu^X\left(k^X(x(t)-\theta) + kx(t) + (k^{XY}-k)y(t) \right) + \sqrt{2\mu^X T^X}\xi^X(t), \\
			\dot{y}(t) &= -\mu^Y\left((k^{YX}-k) x(t) + (k^Y +k)y(t)\right) + \sqrt{2\mu^Y T^Y}\xi^Y(t).
		\end{aligned}
	\end{equation}
	In this setup, the measured system $X$ is trapped in a parabolic potential with stiffness $k^X$ and a minimum at position $\theta$, which is the parameter under consideration, and the measurement system $Y$, which corresponds to the sensor, is trapped in a parabolic potential with stiffness $k^Y$ and a minimum at position 0. They interact with each other via reciprocal interaction $k$ and non-reciprocal interactions $k^{XY}$ and $k^{YX}$. $X$ and $Y$ are attached to heat baths at temperatures $T^X$ and $T^Y$.
	
	As shown in the main text, enhancements in parameter estimation via sensor measurements require non-equilibrium states with non-vanishing probability currents. In particular, non-equilibrium steady states are realized by a temperature difference between subsystems or non-reciprocity in the drift term. The steady-state covariance matrix $\bm{\Xi}^{\rm ss}$, defined by
	\begin{equation}
		\bm{\Xi}^{\rm ss} = \begin{pmatrix}
			\av{(x-\av{x}_{\rm ss})^2}_{\rm ss} & \av{(x y-\av{x y}_{\rm ss})^2}_{\rm ss}\\
			\av{(x y-\av{x y}_{\rm ss})^2}_{\rm ss} & \av{(y-\av{y}_{\rm ss})^2}_{\rm ss}
		\end{pmatrix},
	\end{equation}
	is obtained as the solution to the continuous-time Lyapunov equation (see also \ref{sec-lyapunov}):
	\begin{equation}\label{2d-lyapunov}
		\bm 0 = \bm A \bm{\Xi}^{\rm ss} + \bm{\Xi}^{\rm ss} \bm A^\top + \bm D, 
	\end{equation}
	where
	\begin{equation}
		\bm A = \begin{pmatrix}
			-\mu^X (k^X + k) & -\mu^X (k^{XY} -k)\\
			-\mu^Y (k^{YX}-k) & -\mu^Y(k^Y + k)
		\end{pmatrix} ~\text{and}~ \bm D = \begin{pmatrix}
			2\mu^X T^X & 0\\
			0 & 2\mu^Y T^Y
		\end{pmatrix}.
	\end{equation}
	To ensure the stability of the steady state, we assume that $\bm A$ is negative-definite. The Fisher information of each subsystem at the steady state is expressed as follows (see also \ref{sec-lin-fisher}):
	\begin{equation}\label{fisher-2d}
		F^{X, \rm ss}_\theta = \frac{(\partial_\theta \av{x}_{\rm ss})^2}{(\bm{\Xi}^{\rm ss})_{xx}}, \, F^{Y, \rm ss}_\theta = \frac{(\partial_\theta\av{y}_{\rm ss})^2}{(\bm{\Xi}^{\rm ss})_{yy}},
	\end{equation}
	where $(\bm{\Xi}^{\rm ss})_{xx}$ is defined as the $xx$-component of $\bm\Xi^{\rm ss}$ and $(\bm{\Xi}^{\rm ss})_{yy}$ is defined similarly. From \eqref{2d-lyapunov} and \eqref{fisher-2d}, we obtain the Fisher information ratio in steady states:
	\begin{equation}\label{2d-eta-gen}
		\eta^{\mathrm{ss}} = \frac{F^{Y,\rm ss}_\theta}{F^{X, \rm ss}_\theta} = \frac{(k^{YX}-k)^2}{(k^Y + k)^2} \frac{(k^X + k)(k^Y + k)\mu^X -(k^{XY}-k)(k^{YX}-k)\mu^X + (k^Y + k)^2\mu^Y + (k^{XY}-k)^2\mu^X\frac{T^Y}{T^X}}{(k^X + k)(k^Y + k)\mu^Y - (k^{XY}-k)(k^{YX}-k)\mu^Y + (k^X + k)^2\mu^X + (k^{YX}-k)^2 \mu^Y\frac{T^X}{T^Y}}\frac{T^X}{T^Y}.
	\end{equation}
	The condition for enhanced parameter estimation through sensor measurements, $F^{X, \rm ss}_\theta < F^{Y, \rm ss}_\theta$, corresponds to $\eta^{\mathrm{ss}} > 1$. Previous studies on sensing (such as \cite{Gieseler2013, Cao2013, Fogliano2021}) have demonstrated that cooling the sensor can improve the precision of measurement. Here, we first consider the case of ideal cooling, where $T^Y = 0$. In the $T^Y/T^X \to 0$ limit, we obtain
	\begin{equation}
		\eta^{\mathrm{ss}} \to 1 + \frac{\mu^X}{\mu^Y}\frac{\det\bm A}{((\bm A)_{22})^2} =: \eta^{\mathrm{ss}}_0,
	\end{equation}
	where we denote $(\bm A)_{22} = -\mu^Y(k^Y+k)$ as the 22-component of the drift matrix $\bm A$. From the negative-definiteness of $\bm A$, we have $\det\bm A > 0$, and thus $\eta^{\mathrm{ss}}_0 > 1$. This implies that ideal sensor cooling ($T^Y = 0$) always yields $F^{X, \rm ss}_\theta < F^{Y, \rm ss}_\theta$, i.e., enhanced precision through sensor measurements.
	
\begin{figure}[htbp]
		\centering
		\includegraphics[width=\linewidth]{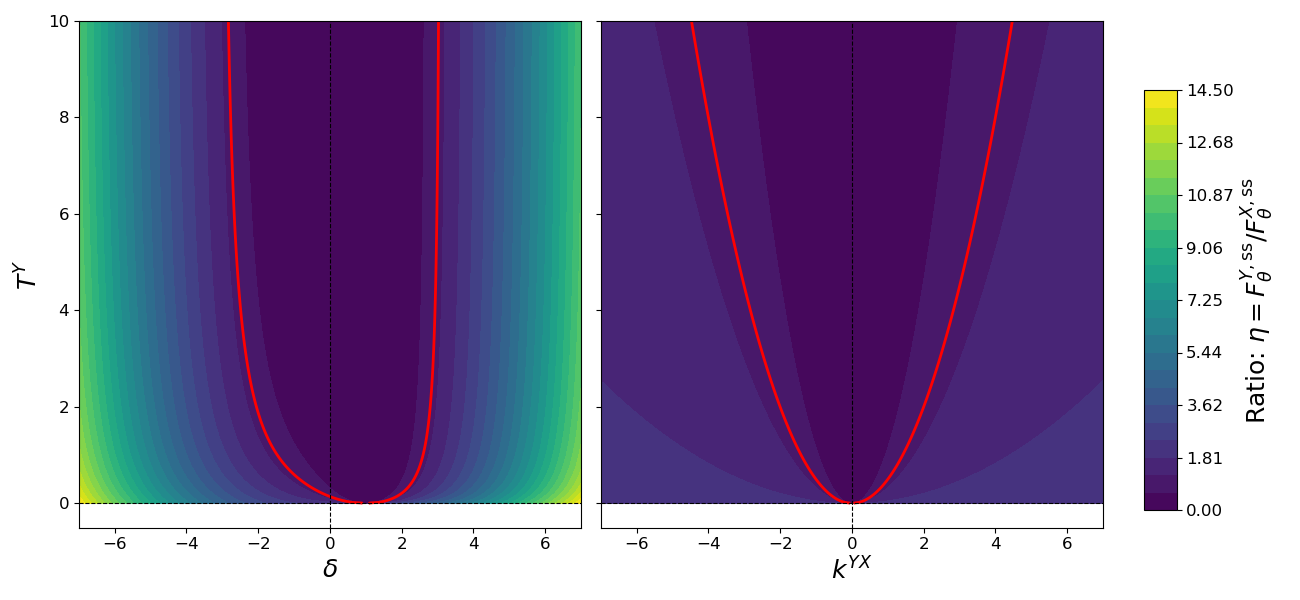}
		\caption{The left panel shows the phase diagram for \eqref{eta-symmetric}. Here, we set $k = 1$ and $T^X = 1$. The remaining parameters are $\mu^X = \mu^Y = 1$. The red line denotes the contour for $\eta^{\mathrm{ss}} = 1$. We observe that the region for $F^{X, \rm ss}_\theta \leq F^{Y, \rm ss}_\theta$ monotonically expands as $T^Y$ decreases. The right panel shows the phase diagram for back-action-free measurements (\eqref{eta-backactionfree}). We set $k^X = k^Y = 1$ and $T^X = 1$. The remaining parameters are $\mu^X = \mu^Y = 1$. The red line denotes the contour for $\eta^{\mathrm{ss}} = 1$. As in the case of \eqref{2d-lin-model-simp} (left panel), the region for $F^{X, \rm ss}_\theta \leq F^{Y, \rm ss}_\theta$ monotonically expands as $T^Y$ decreases.}
		\label{fig:phase}
	\end{figure}

	As mentioned in the previous paragraph and the main text, the non-reciprocity of the drift term is also an important factor for enhanced measurements in addition to the sensor temperature $T^Y$. To investigate the dependence of these factors, we simplify the original dynamics \eqref{2d-lin-model-2}:
	\begin{equation}\label{2d-lin-model-simp}
		\begin{aligned}
			\dot{x}(t) &= -\mu^X\left(k^X(x(t)-\theta) + kx(t) + (-\delta-k)y(t) \right) + \sqrt{2\mu^X T^X}\xi^X(t), \\
			\dot{y}(t) &= -\mu^Y\left((\delta-k) x(t) + (k^Y +k)y(t)\right) + \sqrt{2\mu^Y T^Y}\xi^Y(t).
		\end{aligned}
	\end{equation}
	In \eqref{2d-lin-model-simp}, the non-reciprocity terms $k^{XY}$ and $k^{YX}$ are normalized as $-k^{XY} = k^{YX} = \delta$. Although this system does not fall within the back-action-free measurements, we introduce such a system to examine the behavior of Fisher information in more general systems. $\eta^{\mathrm{ss}}$ in \eqref{2d-eta-gen} is then given by
	\begin{equation}\label{eta-symmetric}
		\eta^{\mathrm{ss}} = \left(\frac{-k + \delta}{k + k^Y}\right)^2\frac{\left((k+k^X)(k+k^Y)-(k-\delta^2)\right)\mu^X + (k+k^Y)^2\mu^Y + (k+\delta)^2\mu^X\frac{T^Y}{T^X}}{\left((k+k^X)(k+k^Y)-(k-\delta^2)\right)\mu^Y + (k+k^X)^2\mu^X + (k-\delta)^2\mu^Y\frac{T^X}{T^Y}}\frac{T^X}{T^Y}.
	\end{equation}
	The left panel of Fig.~\ref{fig:phase} shows the phase diagram for \eqref{eta-symmetric}, where we can see that the region corresponding to $F^{X, \rm ss}_\theta \leq F^{Y, \rm ss}_\theta$ monotonically expands as $T^Y$ decreases. 
	
	Next, we consider back-action-free measurements:
	\begin{equation}\label{2d-lin-model-BA-free}
		\begin{aligned}
			\dot{x}(t) &= -\mu^Xk^X(x(t)-\theta) + \sqrt{2\mu^X T^X}\xi^X(t), \\
			\dot{y}(t) &= -\mu^Y\left(k^{YX} x(t) + k^Y y(t)\right) + \sqrt{2\mu^Y T^Y}\xi^Y(t).
		\end{aligned}
	\end{equation}
	In this case, $\eta^{\mathrm{ss}}$ in \eqref{2d-eta-gen} is expressed as
	\begin{equation}\label{eta-backactionfree}
		\eta^{\mathrm{ss}} = \frac{(k^{YX})^2}{(k^Y)^2}\frac{\mu^Xk^Xk^Y + \mu^Y(k^{Y})^2}{\mu^Y k^X k^Y + \mu^X(k^X)^2 + \mu^Y (k^{YX})^2\frac{T^X}{T^Y}}\frac{T^X}{T^Y}.
	\end{equation}
	The right panel of Fig.~\ref{fig:phase} shows the phase diagram for \eqref{eta-backactionfree}. As in the case of \eqref{2d-lin-model-simp} (left panel), the region corresponding to $F^{X, \rm ss}_\theta < F^{Y, \rm ss}_\theta$ monotonically expands as $T^Y$ decreases.

	\section{Underdamped regime}
	
	In the main text, we have focused only on overdamped systems. In \ref{ud-gen}, we consider measurements processes in underdamped systems and derive the useful formula \eqref{fisher-dt-single}, which is the counterpart to Eq.~(23) in the main text. And in \ref{ud-eq} and \ref{ud-sl}, we extend the results obtained in Appendix B.
	\subsection{Measurements in underdamped regime}\label{ud-gen}
	We consider an underdamped system given by the following equations: 
	\begin{align}\label{underdamp}
		\dot{\bm{x}}(t) &= \bm{v}(t),\nn
		m\dot{\bm{v}}(t) &= -\bm\gamma \bm{v}(t) + \bm{f}(\bm{x}(t)) + \sqrt{2\bm\gamma T}\bm{\xi}(t), 
	\end{align}
	where the dot $\dot{\quad}$ denotes the time derivative, $\bm{\xi}$ represents Gaussian white noise satisfying $\langle\xi_i(t)\rangle=0$ and $\langle\xi_i(t)\xi_j(t')\rangle = \delta_{ij}\delta(t-t')$, and $\bm \gamma$ is the friction tensor. The time evolution of the probability density $p(\bm{x}, \bm{v}, t)$ is governed by the corresponding Fokker-Planck equation \cite{Risken1986, Gardiner2002}:
	\begin{equation}\label{FPE}
		\begin{aligned}
			\dot{p}(\bm{x}, \bm{v}, t) &= \mathcal{L}(\bm{x}, \bm{v})p(\bm{x}, \bm{v}, t), \\
			\mathcal{L}(\bm{x}, \bm{v}) &= -\bm{v}\cdot\nabla_{\bm{x}}-\frac{1}{m}\nabla_{\bm v}\cdot\left[-\bm\gamma\bm v + \bm f (\bm{x}) -\frac{\bm\gamma T}{m}\nabla_{\bm v}\right].
		\end{aligned}
	\end{equation}
	
	Here, we assume that only the drift term $\bm f$ depends on a target parameter $\theta$, denoted as $\bm f_\theta (\bm x)$. We define the Fisher information with respect to $\theta$ as
	\begin{equation}
		F_{\theta}(t) := \int\int d\bm x d\bm v \left(\partial_{\theta}\ln p_{\theta}(\bm x, \bm v, t)\right)^2 p_{\theta}(\bm x, \bm v, t) = \langle\left(\partial_{\theta}\ln p_{\theta}(\bm x, \bm v, t)\right)^2\rangle.
	\end{equation}
	If $\theta$ is multidimensional (i.e., $\bm\theta = \{\theta_i\}_{i = 1}^N$), the Fisher information matrix is defined by its components
	\begin{equation}
		F_{\theta_i\theta_j}(t)= \langle(\partial_{\theta_i}\ln p_{\bm\theta}(\bm x, \bm v))(\partial_{\theta_j}\ln p_{\bm\theta}(\bm x, \bm v))\rangle.
	\end{equation} 
	Hereafter, we assume that $\theta$ is one-dimensional for simplicity. For an unbiased estimator of $\theta$ satisfying $\langle\hat{\theta}\rangle = \theta$, we can derive the Cram\'er-Rao inequality:
	\begin{equation}\label{fisher-component}
		\mathrm{Var}(\hat{\theta}) \geq \frac{1}{F_{\theta}(t)}.
	\end{equation}
	This inequality implies that a larger Fisher information allows for a smaller variance of the estimator, meaning that more accurate parameter estimation is possible. That is, the Fisher information quantifies how informative the system's observations are about the parameter.
	
	The time derivative of $F_{\theta}(t)$ is calculated as follows:
	\begin{align}
		\dot{F}_{\theta}(t) &= \int\int d\bm x d\bm v \left(\partial_{\theta}\ln p_{\theta}(t)\right)^2 \dot{p}_{\theta}(t) + \int\int d\bm x d\bm v \frac{d}{dt}\left[\left(\partial_\theta\ln p_{\theta}(t)\right)^2\right] p_{\theta}(t),\nn
		&= \int\int d\bm x d\bm v \left(\partial_{\theta}\ln p_{\theta}(t)\right)^2 \dot{p}_{\theta}(t) + 2\int\int d\bm x d\bm v \left(\partial^2_{t, \theta}\ln p_{\theta}(t)\right)\left(\partial_\theta\ln p_\theta(t)\right)p_\theta(t).
	\end{align}
	For overdamped dynamics,
	\begin{align}
		\dot{\bm x}(t) = \bm\mu \bm f_\theta(\bm x(t)) + \sqrt{2\bm D}\bm\xi(t),
	\end{align}
	we have already derived the relation (Eq.~(23) in Appendix B).
	\begin{equation}
		\dot{F}_\theta(t) = -2\av{\partial_\theta \grad_x \ln p_\theta(t) \cdot \bm D\partial_\theta\grad_x\ln p_\theta(t)} + 2\av{\partial_\theta \grad_x \ln p_\theta(t) \cdot\partial_\theta (\bm\mu\bm f_\theta)},
	\end{equation}
	where $\bm{\mu}$ is the mobility matrix and $\bm{D}$ is the diffusion matrix. By treating the underdamped system \eqref{underdamp} as an overdamped system, we can utilize the above relation: The dynamics \eqref{underdamp} is reformulated as
	\begin{align}
		\dot{\bm X}(t) = \bm{A}_\theta(\bm X(t)) + \sqrt{2\bm B}\bm\zeta(t),
	\end{align}
	where $\bm X = (\bm x, \bm v)^\top$ and $\bm\zeta(t) = (\bm \xi^x(t), \bm\xi(t))^\top$. Here, $\bm\xi^x(t)$ is defined as Gaussian white noise satisfying $\av{\xi^x_i}=0$ and $\av{\xi_i^x(t)\xi^x_j(t')} = \delta_{ij}\delta(t-t')$. Also, we have defined the drift vector $\bm A_\theta$ as
	\begin{equation}
		\bm A_\theta = \begin{pmatrix}
			\bm v(t)\\
			-\frac{\bm \gamma}{m}\bm v(t) + \frac{\bm f_\theta(\bm x(t))}{m}
		\end{pmatrix},
	\end{equation}
	and the diffusion matrix $\bm B$ as
	\begin{equation}
		\bm B = \begin{pmatrix}
			\bm 0 & \bm 0\\
			\bm 0 & \frac{\bm \gamma T}{m^2}
		\end{pmatrix}.
	\end{equation}
	The time derivative of the Fisher information is then obtained as
	\begin{equation}\label{fisher-dt-single}
		\dot{F}_\theta(t) = -2\av{\partial_\theta \grad_X \ln p_\theta(t) \cdot \bm B\partial_\theta\grad_X\ln p_\theta(t)} + 2\av{\partial_\theta \grad_X \ln p_\theta(t) \cdot\partial_\theta (\bm A_\theta)}.
	\end{equation}
	
	Now, let us consider measurement processes in underdamped systems. As in the case of the overdamped system (main text), we express the measurement process using bipartite systems:
	\begin{align}\label{UD-bip}
		\dot{\bm{x}} &= \bm v, \quad m^X\dot{\bm{v}} = -\bm\gamma^X \bm{v} + \bm{f}^X_\theta(\bm{x}, \bm{y}) + \sqrt{2\bm\gamma^X T^X}\bm{\xi}^X(t),\nn
		\dot{\bm{y}} &= \bm w, \quad m^Y\dot{\bm{w}} = -\bm\gamma^Y \bm{w} + \bm{f}^Y(\bm{x}, \bm{y}) + \sqrt{2\bm\gamma^Y T^Y}\bm{\xi}^Y(t).
	\end{align}
	Hereafter, we denote $X$ as the measured system and $Y$ as the measurement system (sensor). The Cram\'er-Rao inequality \eqref{fisher-component} gives a lower bound for the precision of the measurement; a higher value of the Fisher information implies a more precise measurement (i.e. lower $\operatorname{Var}(\hat \theta)$). 
	
	\subsection{Equilibrium measurements}\label{ud-eq}
	To begin with, we consider measurements in the equilibrium regime. We assume that the drift term is driven by a conservative force;
	\begin{equation}\label{potential}
		\begin{pmatrix}
			\bm f_\theta^X(\bm x, \bm y)\\
			\bm f^Y(\bm x, \bm y)
		\end{pmatrix}=-\grad_{xy}U_\theta^{XY}(\bm x, \bm y) = -\grad_{xy}U_\theta^X(\bm x) - \grad_{xy}U^Y(\bm y) - \grad_{xy}U^{\text{int}}(\bm x, \bm y),
	\end{equation}
	and $T^X = T^Y = T$ holds. In this case, the steady state is an equilibrium state with vanishing probability current \cite{shiraishiIntroductionStochasticThermodynamics2023a}:
	\begin{equation}\label{eq-prob}
		p_\theta^{\stst}(\bm x, \bm y, \bm v, \bm w) = p_\theta^{\eq}(\bm x, \bm y, \bm v, \bm w) \propto \exp\left[-\beta\left(\frac{m^X\bm v^2}{2}+\frac{m^Y\bm w^2}{2} + U_\theta^{XY}(\bm x, \bm y)\right)\right],
	\end{equation}
	where $\beta = 1/T$. In the same way as in the overdamped case, we can show that
	\begin{equation}\label{fisher-equilibrium-underdamp}
		F_\theta^{Y, \eq} \leq F_\theta^{X, \eq}
	\end{equation}
	holds as follows: From \eqref{eq-prob}, we obtain
	\begin{align}
		p^{Y \vert X,\text{eq}}_\theta(\bm{y} \vert \bm{x}) = \frac{\exp\big[-\frac{U^Y(\bm{y}) + U^\text{int}(\bm{x},\bm{y})}{T}\big]}{\int d\bm{y} \ \exp\big[-\frac{U^Y(\bm{y}) + U^\text{int}(\bm{x},\bm{y})}{T}\big]}.
	\end{align}
	Since the expression above is independent of $\theta$, we attain $F_\theta^{Y \vert X,\text{eq}} = 0$, and thus, from the chain rule of Fisher information,
	\begin{align}
		F_\theta^{XY,\text{eq}} = F_\theta^{Y \vert X,\text{eq}} + F_\theta^{X,\text{eq}},
	\end{align}
	$F_\theta^{XY,\text{eq}} = F_\theta^{X,\text{eq}}$ holds. This means that, in equilibrium,  further measurements of $Y$ do not provide enhanced parameter estimations on $\theta$.
	
	On the other hand, the chain rule also gives
	\begin{align}
		F_\theta^{XY,\text{eq}} = F_\theta^{X \vert Y,\text{eq}} + F_\theta^{Y,\text{eq}}.
	\end{align}
	In this case, since the probability of $X$ conditioned on $Y$ depends on $\theta$ and is  given by
	\begin{align}
		p^{X \vert Y,\text{eq}}_\theta(\bm{x} \vert \bm{y}) = \frac{\exp\left[-\frac{U^X_\theta(\bm{x}) + U^\text{int}(\bm{x},\bm{y})}{T}\right]}{\int d\bm{x} \exp\left[-\frac{U^X_\theta(\bm{x}) + U^\text{int}(\bm{x},\bm{y})}{T}\right]},
	\end{align}
	$F_\theta^{X \vert Y,\text{eq}} \geq 0$ holds. This implies $F_\theta^{Y,\text{eq}} \leq F_\theta^{XY,\text{eq}}$ and, thus, $F_\theta^{Y,\text{eq}} \leq F_\theta^{X,\text{eq}}$ in equilibrium, which is \eqref{fisher-equilibrium-underdamp}.
	
	\subsection{Speed limit and its derivation}\label{ud-sl}
	Now we derive the speed limit relation for measurement processes in the underdamped regime. The time derivative of Fisher information for the bipartite system \eqref{UD-bip} is explicitly written as
	\begin{align}
		d_t F^{XY}_\theta(t) = -2&\Av{\partial_\theta\grad_v\ln p_\theta^{XY} \cdot\frac{\bm\gamma^X T^X}{(m^X)^2}\partial_\theta\grad_v\ln p_\theta^{XY}}-2\Av{\partial_\theta\grad_w\ln p_\theta^{XY}\cdot\frac{\bm\gamma^Y T^Y}{(m^Y)^2}\partial_\theta\grad_w\ln p_\theta^{XY}} \nn
		&+2\Av{\partial_\theta \grad_v\ln p_\theta^{XY}\cdot\frac{1}{m^X}\partial_\theta\bm f^X_\theta}.
	\end{align}
	By defining $\mathfrak{B}^X_\theta(t)$ and $\mathfrak{F}^X_\theta(t)$ as
	\begin{equation}
		\mathfrak{B}^X_\theta(t) = \Av{\partial_\theta\grad_v\ln p_\theta^{XY} \cdot\frac{\bm\gamma^X T^X}{(m^X)^2}\partial_\theta\grad_v\ln p_\theta^{XY}}, \quad \mathfrak{F}^X_\theta(t)=\frac{1}{T^X}\Av{\partial_\theta \bm f_\theta^X(\bm x)\cdot\bm\gamma^{X, -1}\partial_\theta\bm f_\theta^X(\bm x)},
	\end{equation}
	and utilizing the Cauchy-Schwarz inequality, we arrive at
	\begin{align}
		d_t F^{XY}_\theta(t) \leq -2\mathfrak{B}^X_\theta(t)+2\sqrt{\mathfrak{B}^X_\theta(t)\mathfrak{F}^X_\theta(t)}.
	\end{align}
	Maximizing the right-hand side with respect to $\sqrt{\mathfrak{B}^X_\theta(t)}$ yields
	\begin{equation}
		d_t F^{XY}_\theta(t) \leq \frac{1}{2}\mathfrak{F}^X_\theta(t),
	\end{equation}
	which can be regarded as the speed limit relation. Also, if $X$ and $Y$ are decoupled at $t=0$ and the measurement is back-action-free, we obtain an upper bound on $F_\theta^Y(t)$:
	\begin{equation}\label{speed-limit-underdamped}
		F_\theta^Y(t)\leq F^{X,\eq}_\theta + \frac{1}{2}\int_0^t dt'\, \mathfrak{F}_\theta^X(t'),
	\end{equation}
	which is the counterpart of Eq.~(6) in the main text. When $X$ is linear and one-dimensional expressed as 
	\begin{equation}
		m^X\dot{v} = -\gamma^Xv - k^X(x-\theta) + \sqrt{2\gamma^XT^X}\xi^X,
	\end{equation}
	$\mathfrak{F}_\theta^X(t)$ is explicitly given by
	\begin{equation}
		\mathfrak{F}^X_\theta(t) = \frac{(k^X)^2}{\gamma^X T^X} = \text{const.}
	\end{equation}
	
	\subsection{Numerical demonstrations}
\begin{figure}[htbp]
		\centering
			\includegraphics[width=0.47\textwidth]{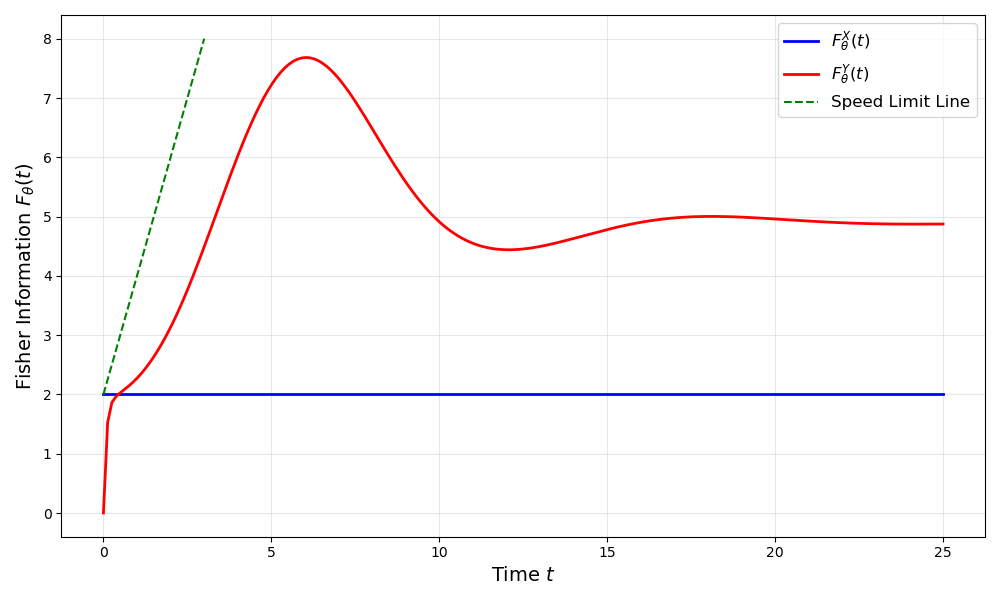}
		\hfill
			\includegraphics[width=0.47\linewidth]{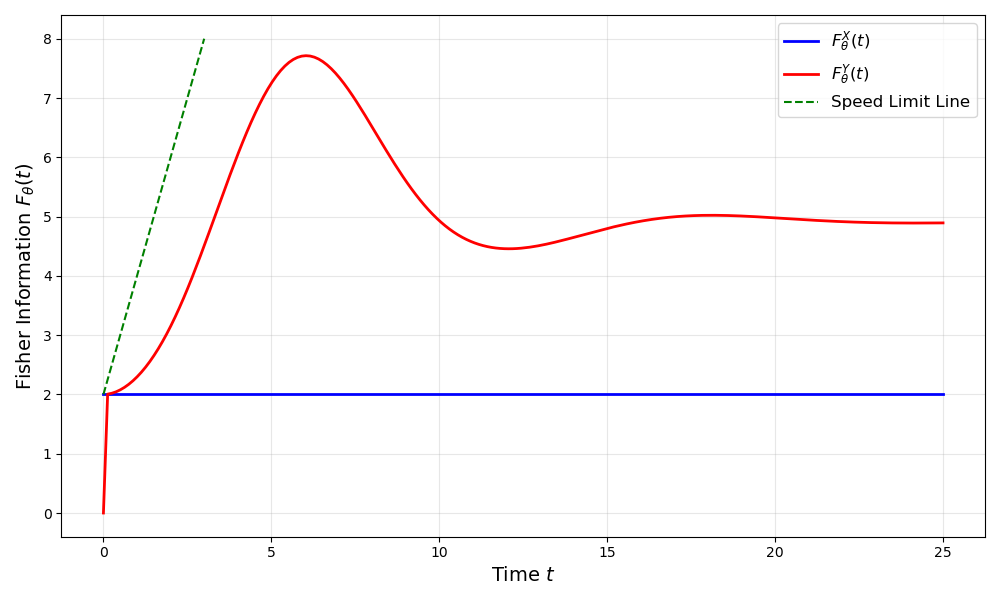}
		\caption{Fisher information dynamics for an underdamped system. The measurement system $Y$ conducts back-action-free measurement on the measured system $X$: $\kappa = k^{XY}=0$ and $k^{YX} = 12$ (the measurement is back-action-free). In the left panel, we assume that the composite system is attached to a single heat bath ($T^X = T^Y=1$). The friction coefficients of the system and sensor are different: $\gamma^X = 1.0, \gamma^Y = 1.5$. The remaining parameters are $k^X = 2, k^Y= 1$, and $m^X = 1, m^Y = 3$. The red line denotes the time evolution of the Fisher information of measurement system $F_\theta^X(t)$, and the blue line denotes that of measured system $F^Y_\theta(t)$. And the dashed green line denotes the speed limit line \eqref{speed-limit-underdamped}, which upper bounds $F^Y_\theta(t)$. The enhanced precision though sensor measurement is realized; $F^{X, \rm ss}_\theta<F^{Y, \rm ss}_\theta$. The right panel shows the noiseless measurement of an underdamped system with an ideally cooled sensor ($T^Y = 0$). The remaining parameters are the same as in the left panel.}
		\label{fig:underdamp_all}
	\end{figure}

	We numerically investigated a two-dimensional linear system (system of interest $X$: 1D, sensor $Y$: 1D):
	\begin{align}
		\dot{x} &= v, \quad m^X\dot{v} = -\gamma^X v -k^X(x-\theta) - \kappa(x-y) - k^{XY}y + \sqrt{2\gamma^X T^X}\xi^X(t),\nn
		\dot{y} &= w, \quad m^Y\dot{w} = - \gamma^Y w -k^Y y - \kappa(y-x) - k^{YX}x + \sqrt{2\gamma^Y T^Y}\xi^Y(t).
	\end{align}
	Fig.~\ref{fig:underdamp_all} shows the results for a backaction-free measurement ($\kappa = k^{XY}=0$). With these parameters, we observe $F^{X,\rm ss}_\theta \leq F^{Y, \rm ss}_\theta$, which implies that precision is enhanced in the steady-state measurement on the sensor $Y$.
	
\end{document}